# Graph Neural Networks at War: Integrating Cybersecurity and Drone Intelligence in the Israeli-Iranian Conflict

*Sozan Sulaiman Maghdid*[1*], *Tarik Ahmed Rashid*[2], *Shavan Askar*[3]

[1]Department of Information Technology, Khabat Technical Institute; Erbil Polytechnic University, Erbil, Iraq

*Corresponding email: sozan.maghdid@epu.edu.iq

[2]Computer Science and Engineering Department; AIIC, University of Kurdistan Hewler, Erbil, Iraq

tarik.ahmed@ukh.edu.krd

[3]Department of Information Systems Engineering, Technical College of Computer and Informatics Engineering, Erbil Polytechnic University, Erbil, Iraq

shavan.askar@epu.edu.iq

**Abstract**

Physical cyber systems have brought about new threats and challenges in detection and immediate response. This study examines how Graph Neural Networks (GNNs) can be used to aid cybersecurity and drone management in a physical cyber system comprising of cyber intrusions and unmanned aerial vehicles (UAVs). By providing a bridge between structural understanding of graphical neural networks, this work has provided an integrated procedure that allows intrusion detection systems to educate on underlying network structures, identify malicious activity, and facilitates drone response measures. Based on an emulation-based case study, cyberattacks models were created to provoke the responses of the drones, which proved that graph-based learning can assist with the situational awareness, swarm coordination, and adaptive maneuver. According to the performance valuation, this method has a detection rate of 94.2, average area under the receiver operating characteristic (ROC) of 0.955 and an average response time of 1.4 seconds. Comparative experiments reveal that proposed GraphSAGE network is more effective than the Graphical Convolutional Networks (GCNs) and Graphical Attention Networks (GATs) in the identical situation. Such findings prove that graphical neural networks can be used to avert intrusion and response of dynamic cyber-physical systems.

## 1. Introduction

The landscape of modern warfare is undergoing a radical transformation. Traditional battlefields have moved beyond physical terrain to include cyberspace, where conflicts are now fought using code, algorithms, and autonomous systems [1]. In this era of hybrid warfare, the integration of cyber-attacks with physical assets, such as drones and autonomous vehicles, is increasingly common, posing a new set of challenges for national defence systems [2]. This dual-domain conflict environment requires innovative solutions capable of operating across both the cyber and physical domains simultaneously [3]. The Israeli Iranian conflict exemplifies this development. Both countries have engaged in high-profile cyber operations, including the Stuxnet worm targeting Iranian nuclear facilities [4], Israeli cyberattacks on Iranian infrastructure [5], and retaliatory Iranian cyberattacks on vital Israeli systems [6]. At the same time, both sides have intensified the use of drones for surveillance, reconnaissance, and targeted strikes. These drones operate in contested environments, deprived of GPS, making them vulnerable to jamming, spoofing, and cyber-physical attacks [7].

Despite advances in both cybersecurity and drone technologies, these systems are often developed in isolation. Intrusion detection systems (IDSs) typically use rule-based or statistical anomaly detection methods, which cannot generalise to unseen threats or account for structural dependencies in networks [8]. Similarly, drone swarm intelligence is primarily designed to optimise spatial trajectory or target tracking, with little or no integration of real-time cyber threat intelligence. This disconnect poses a serious limitation for mission-critical systems operating under coordinated cyber-physical threats [9]. This isolated approach is fundamentally inadequate for modern military applications, where cyberattacks can directly impact physical behaviour (e.g., jamming drone signals, GPS spoofing, or command injection), and where physical responses may need to be triggered automatically through cyber threat detection. There is an urgent need for unified systems capable of interpreting cyber-physical interactions, detecting anomalies, and coordinating response actions across both domains [10]. Graph neural networks offer an effective solution to this challenge. Unlike traditional machine learning models, these networks are designed to learn from graphically structured data by modelling the properties of nodes and the relationships between them [11] . This capability makes them ideal for cybersecurity applications [12], communications network [13] , and data swarm coordination [14].

One of the most important advantages of GNNs in cyber-physical defence systems is their scalability and adaptability. Models such as Graph Convolutional Networks, Graph Attention Networks, and Graph SAGE enable local message-passing computations that can adapt to instantaneous changes in graph structure [15]. This is essential for military environments characterised by dynamic topology, such as swarms of drones adjusting their formations or mobile sensor networks navigating hostile terrain. Furthermore, GNNs can operate in inductive environments, enabling learned models to be generalised to previously unseen graph instances, a critical capability in unpredictable or hostile battlefield conditions [16]. In the field of cybersecurity, GNNs have proven exceptionally effective. They enable protocol-independent intrusion detection by learning patterns of natural system behaviour and identifying anomalies at the graph level, without the need for prior decoding of message formats or specific attack signatures [17]. This enables early detection of message spoofing, command injection, or signal manipulation targeting critical components such as the control area network (CAN) bus or unmanned aerial vehicles (UAVCAN) in autonomous platforms. As cyber-physical military assets increasingly operate beyond the line of sight (BVLOS) and rely on network communications, the ability to secure critical control paths in real-time has become a strategic necessity [18].

Moreover, GNNs assist in creating autonomous decision making agents in terms of graphical representations of the battlefield [19]. As an example, situational awareness graphs (e.g., threat, allies, resources, and terrain) are modeled with the help of tactical edge devices and GNN-based policies are trained through reinforcement learning and inter-agent cooperation. Such applications broaden the boundaries of AI-driven mission autonomy, minimizing the decision-making time, and maximizing survivability in contested space [20]. The GNNs are a core element of the AI architecture of military cyber-physical systems, which delivers the mathematical formulation, computation efficiency, and scale of highly confident defensive operations [21]. GNNs will play a central role in facilitating intelligent coordination, strong security, and strategic advantage in various fronts, air, land, sea, cyber, and space, as data-driven and autonomous conflicts become more and more commonplace [22].

In the context of the Israeli-Iranian hostilities, graph neural networks have emerged as a vital enabler for enhancing the cybersecurity and operational autonomy of modern cyber-physical military systems, especially unmanned aerial vehicles (UAVs). UAVs now play a critical role in intelligence gathering, precision strikes, and electronic warfare, but their increasing reliance on lightweight communication protocols such as the Controller Area Network (CAN) bus and its UAV-focused variant, UAVCAN, exposes them to significant cyber vulnerabilities [23]. These protocols, while cost-effective and efficient for real-time intra-system communication, were never designed with adversarial threats in mind and lack fundamental security mechanisms like encryption, authentication, or message integrity verification. This makes them susceptible to sophisticated attacks such as spoofing, malicious message injection, and remote command hijacking, threats that are no longer theoretical but have been demonstrated in high-profile incidents where drones have been intercepted or manipulated in conflict zones [24].

This paper proposes an innovative and integrated graph neural network -based framework that tightly combines intrusion detection and autonomous drone coordination within a single unified architecture. Our framework enables intrusion detection systems to learn from the topological and behavioural properties of networked systems, accurately detect malicious activities, and launch real-time countermeasures against drones. The drone component of the system acts as a GNN-inspired swarm, capable of adapting flight paths, organising cooperative defences, and performing evasive manoeuvres based on the evolving threat landscape. To validate our approach, we developed a simulation case study based on the Israeli Iranian cyber-drone confrontation. In this scenario, a GNN detects a series of cyberattacks, simulating node intrusions and link outages, which in turn trigger a response based on drone swarms. Each drone acts as an intelligent agent in the graph, updating its position and mission through learned GNN policies. This simulation demonstrates how our system enhances situational awareness, automatic threat neutralisation, and decision-making time in a contested environment.

The system is evaluated across four key dimensions: (i) *Intrusion Detection Accuracy*, which measures the system's ability to correctly classify healthy versus compromised nodes; (ii) *Threat Neutralisation Rate*, referring to the proportion of detected threats successfully intercepted or neutralised by the drones; (iii) *Response Time*, indicating the latency between threat detection and the initiation of countermeasures; and (iv) *Swarm Efficiency and Cohesion*, which assesses the drones' ability to maintain coordination during evasion and engagement operations.

This work positions graph neural networks as a strategic tool in multi-domain warfare and advocates for their use in critical, real-time defence applications. By connecting the cyber and physical domains through graph-based intelligent learning, we take a critical step toward autonomous and integrated national security systems suited to the realities of 21st-century conflict.

## 1.1 Problem declaration

Modern warfare is increasingly defined by the integration of cyber and physical domains, creating complex threat environments that demand intelligent, adaptive, and real-time defense mechanisms. This convergence introduces a series of critical challenges that current systems are ill-equipped to handle. One of the foremost challenges is the hybrid nature of modern threats. Contemporary battlefields—such as those observed in the Israel–Iran conflict—are shaped by a fusion of cyber and physical attack vectors. These include advanced persistent threats (APTs), GPS spoofing, communication jamming, and unauthorised UAV incursions. Such threats place both autonomous drone operations and critical military infrastructure at significant risk, requiring defense systems that can respond fluidly across domains. Existing intrusion detection systems (IDSs) are inadequate in this regard. Predominantly rule-based or packet-centric, traditional IDSs are inherently inflexible and fail to capture the relational complexity and dynamic behavior patterns that characterize cyber-physical warfare. Their limited adaptability reduces their effectiveness in identifying novel or evolving attack strategies.

Moreover, current cybersecurity solutions often operate in isolation from physical response mechanisms, resulting in fragmented situational awareness and delayed threat mitigation. This lack of integration significantly impairs coordinated responses during active multi-domain engagements, where time-sensitive decisions are critical. Another major shortcoming lies in the inadequate modeling of dynamic battlefield environments. Conventional algorithms tend to ignore the real-time telemetry of UAVs and the evolving interdependencies among nodes in the network. This oversight undermines the system's ability to maintain situational awareness and execute informed decisions in real time. Addressing these gaps requires the development of a unified analytical framework that integrates cyber and physical threat intelligence. Such a system must be capable of analyzing threats in a holistic cyber-physical context, predicting cascading effects across interconnected components, and initiating autonomous, decentralized, and context-aware drone responses. Without such an integrated and intelligent framework, military operations will remain vulnerable to increasingly sophisticated and fast-evolving multi-domain threats—threats that are advancing at a pace beyond the capabilities of conventional detection and mitigation systems.

## 1.2 Study gap

While numerous studies have explored cybersecurity techniques and intrusion detection systems (IDS), most remain limited to static environments and fail to account for the dynamic interaction between cyber and physical systems in modern warfare scenarios. Current cyber threat intelligence models often focus exclusively on extracting indicators of compromise or identifying threat actors, lacking the contextual awareness needed to interpret attack progression across interconnected systems. Moreover, current applications of graph neural networks in cybersecurity primarily target enterprise networks or social graphs, with limited exploration of critical, defence-oriented, and immediate applications. There is insufficient research addressing how GNNs can be used to integrate cyber data with physical drone intelligence in high-risk conflict zones, such as the Israel-Iran theatre of operations. Furthermore, few approaches integrate GNN-based intrusion detection with autonomous drone coordination to proactively neutralise threats. The absence of such a unified, intelligent framework limits the ability to detect, analyse, and respond to threats quickly and accurately in cyber-physical conflict environments.

## 1.3 Contribution

This research paper presents the following contributions:

1. A unified framework that integrates cyber and physical systems, combining neural network-based intrusion detection with drone swarm coordination.
2. A graphical representation of drone networks, where telemetry and drone communications data are modeled as dynamic graphs, enabling the application of GraphSAGE for multi-category intrusion detection.
3. A coordination mechanism that links node-level threat predictions with system-level drone responses, allowing for adaptive actions such as rerouting and configuration adjustments.
4. A simulation-based implementation of the proposed framework using a cyber drone dataset that reflects realistic telemetry and attack scenarios.
5. A comprehensive evaluation of the framework using classification metrics (accuracy, fine-tuning, recall, and F1), confusion matrix analysis, ROC-AUC assessment, and system performance indicators such as response time and coordination effectiveness.
6. Comparative analysis of GraphSAGE with GCN and GAT models under consistent experimental conditions.

## 1.4 Research Objectives

This study aims to design an integrated graph neural network-based framework capable of countering real-time cyber and physical threats in multi-domain warfare environments. The research focuses on four specific objectives that together form the foundation of the proposed system.

First, the study seeks to build a dynamic graphical model of the cyber-physical battlefield, representing key elements such as communication nodes, UAVs, and potential threats. This graph-based abstraction provides a flexible and scalable structure for modelling complex interactions in contested environments.

Second, a GNN-driven intrusion detection system will be implemented. This component is designed to learn from both topological and behavioural patterns in the cyber-physical graph to accurately detect cyber-attacks, including sophisticated and evolving threat vectors.

Third, the framework will coordinate the response of autonomous UAVs using decisions informed by the GNN model. This enables swarm-based behaviours for threat neutralisation and evasive action, allowing drones to adapt collectively and autonomously to unfolding attack scenarios.

Finally, the system's performance will be evaluated using a set of key metrics, including intrusion detection accuracy, response time, threat neutralisation rate, and swarm efficiency. These metrics will assess the effectiveness and responsiveness of the framework in simulated multi-domain warfare conditions.

### 1.5 Novelty Statement

This work presents an integrated framework combining graphical neural network-based intrusion detection and drone swarm coordination in a cyber-physical environment. Rather than proposing a new architecture for graphical neural networks, this study's contribution lies in designing and evaluating a unified system that links graphical threat detection with coordinated response mechanisms.

This framework designs drone networks as dynamic graphs and applies the GraphSAGE architecture to monitor the structural and contextual dependencies of multi-category threat classification. The novelty of this work lies in its integration of graph-based learning with system-level adaptive coordination, supported by a comprehensive assessment using classification metrics, confusion matrices, ROC-AUC analysis, and operational performance indicators.

This approach demonstrates how existing graphical neural network models can be effectively applied within cyber-physical systems to support intrusion detection and coordinated drone response in dynamic environments.

## 2. Related Work

The evolving nature of modern conflict has led to an increasing convergence between cyber warfare and autonomous systems, particularly drones. Recent literature reflects growing interest in understanding this intersection, as nations leverage both digital and kinetic capabilities to achieve strategic objectives. To contextualise the need for integrated analytical approaches, we first examine a representative real-world case where cyber operations and drone technologies are deeply intertwined: the Israeli-Iranian conflict.

## 2.1 Israeli-Iranian Conflict in Cyber and Drone Domains

Borg [25] asserted that the Israeli-Iranian conflict represents one of the most prominent and well-documented cases of hybrid warfare, with cyber operations and unmanned aerial systems becoming crucial tools in shaping the strategic dynamics of the Middle East. This development highlights how cyber and physical domains are increasingly intertwined, impacting regional security and redefining traditional warfare models [25]. Eiran [26] confirms that Israel is considered a regional and global leader in drone technology, with the IDF deploying advanced drones extensively for intelligence, surveillance, target acquisition, and reconnaissance purposes. Furthermore, these drones play a pivotal role in directing precision strikes against enemy assets and infrastructure, highlighting Israel's technological superiority in integrating drones into modern military strategies [26]. Stoddart [27] highlights that Israel's cyber capabilities have been pivotal in both offensive and defensive cyber operations, reflecting its prominent role in shaping digital warfare. A prime example of this is the alleged Stuxnet attack on Iran's Natanz nuclear facility, widely believed to have been the product of Israeli-American collaboration. This incident illustrates how advanced malware can be used to disrupt critical infrastructure not only as a tactical tool but also as a strategic deterrent in the evolving landscape of cyber-physical conflict [27]. Barak et al. [28] emphasise that Israel's National Cyber Security Directorate plays a pivotal role in coordinating cyber defence efforts across the civilian and military sectors, ensuring an integrated approach to national security. By combining centralised threat intelligence with offensive cyber capabilities, Israel enhances its ability to anticipate, mitigate, and respond to complex cyber threats, highlighting the country's advanced position in managing hybrid security challenges [28].

Abbas and Fatima [29] indicate that Iran has responded by intensifying its domestic efforts to develop drone technology and cyberwarfare capabilities, using these tools as part of an asymmetric strategy aimed at counterbalancing Israel's technological superiority. This trend reflects Iran's broader goal of expanding its influence and resilience in hybrid conflict environments by developing indigenous drone fleets and advanced cyber units [29]. Borsari highlights that the Islamic Revolutionary Guard Corps (IRGC) maintains an expanding fleet of domestically developed drones, including loitering munitions and long-range reconnaissance UAVs. These assets have been actively deployed across multiple theatres such as Syria, Iraq, and the Persian Gulf, underscoring Iran’s efforts to project power and extend its asymmetric capabilities in regional conflicts [30]. Arshad [31] observed in his study that Iranian cybersecurity units, including APT33 and APT34, are linked to cyber espionage and sabotage campaigns targeting critical Israeli infrastructure. These operations have targeted sectors such as water utilities and private companies, demonstrating Iran's increasing investment in offensive cyber capabilities as part of its broader asymmetric strategy [31] . Recent studies demonstrate how this conflict increasingly intersects cyber and drone operations. For example, Stachoń [32] highlights allegations that Iran has attempted to penetrate Israeli drone command and control systems through sophisticated cyber intrusions. In response, Israeli advanced signals intelligence (SIGINT) and electronic intelligence units are exploiting intercepted drone telemetry data and recovered devices to gather valuable cyber intelligence, deepening the cyber dimension of drones in their ongoing strategic competition [32]. Stachoń [32] also emphasizes that the resulting operational environment is a tightly interconnected cyber-physical battlespace where drones function not only as kinetic assets but also as data nodes in broader intelligence and network operations [32]. Khan [33] argued that despite significant technological advances, current analytical frameworks often treat cyber and drone domains as separate areas, ignoring the emerging interconnections that increasingly define this modern theatre of conflict [33]. This gap underscores the need for integrated models and methods that can integrate multi-domain intelligence streams, uncover hidden connections, and support real-time decision-making, a role for which graph-based learning methods, such as graph neural networks, are increasingly promising.

## 2.2 Graph Neural Networks

Mughal et al. [34] introduced a GNN-based intrusion detection system using a testbed of six UAVs arranged in a hexagonal formation. By simulating various cyberattacks, the study collected spatial and temporal data to assess whether incorporating UAV swarm topology improves threat detection [34]. Majumder et al. [35] proposed a graph-based IDS for UAVs using the UAVCAN protocol by converting CAN messages into graph structures and applying models such as GCN, GAT, and GraphSAGE. The study found that inductive models like GAT and GraphSAGE outperformed LSTM and transductive approaches [35]. Du et al. [36] proposed DEGNN, a GNN-based model for binary code similarity analysis using call and attention-enhanced control graphs. This model generates dual-function embeddings and excels in drone cybersecurity tasks, outperforming existing models with up to a 33.3% improvement in cross-architecture vulnerability detection [36]. Feng et al. [37] proposed CEG (Causality-Enhanced Graph Neural Network), an unsupervised anomaly detection model for UAVs. CEG integrates causal relationships among flight parameters using a graph attention mechanism to extract spatial features and a trend-decomposed module for temporal dependencies. Evaluated on the ALFA dataset, CEG significantly outperforms existing methods in Precision, Recall, and F1 score [37].

Zhang and Li [38] proposed SR-GNN, a self-supervised anomaly detection framework for UAV trajectory data. This framework combines spectral residuals to remove point anomalies and dual GNNs to capture time series relationships. The SR-GNN was jointly optimised through prediction and reconstruction and evaluated on a custom trajectory-adjusted dataset, outperforming existing models in anomaly detection accuracy [38]. Njanko et al. [39] introduced a GNN-based deception system to counter Network Reconnaissance Attacks (NRA) on UAVs. By using a Graph Neural Network to select fake network configurations that mimic the structure of the defender's UAV network but differ in features, the system delivers misleading responses to attackers, enhancing UAV cybersecurity through strategic deception [39]. Akram et al. [40] developed UAVGuard, an attack model targeting vulnerabilities in Graph Neural Networks, particularly label-flipping attacks in UAV systems. UAVGuard uses continuous approximations and a simplified GNN structure to perform effective gradient-based attacks. To defend against this, they proposed a community-preserving self-supervised regularisation framework that enhances model robustness. Evaluations on real-world datasets confirmed both the potency of UAVGuard and the resilience of the proposed defence [40]. Kumar et al. [41] proposed a cybersecurity framework using Cross-Layer Adaptive Graph Neural Networks to detect cyber threats in VANETs, focusing on Roadside Units (RSUs). By modelling the network architecture as a graph and incorporating information from multiple protocol layers, his approach identifies complex attack patterns and adapts in real-time to evolving threats [41]. Du et al.[42] introduced a hybrid intrusion detection approach for UAV networks by first proposing a graph construction method suited for scenarios with few ECU nodes to enhance information density. They then developed UAV-GCNLSTM (UGL), a model combining Graph Convolutional Networks for structural learning with LSTM for temporal pattern recognition [42]. Sun et al. [43] developed a GNN-based Intrusion Detection System (GNN-IDS) that combines attack graphs with real-time network data to model both static and dynamic aspects of computer networks. Leveraging Graph Neural Networks, the system effectively learns node relationships to detect anomalies and pinpoint the malicious actions behind them. Tested on synthetic and public IDS datasets, GNN-IDS demonstrated high performance with strengths in uncertainty management, interpretability, and robustness [43].

Dang et al. [44] proposed a graph neural network-based method for detecting GPS spoofing in cellular-connected UAV swarms. The approach analyses the similarity between swarm GPS and communication topologies to identify spoofing attempts. A UAV swarm simulator, built using bipartite graphs and the Hungarian algorithm, was employed for evaluation [44]. Friji et al. [45] introduced a novel intrusion detection framework that models network communication as a graph of flows, capturing key topological insights such as multi-step attack phases and spoofing activity relationships. The system uses a Graph Neural Network-based classifier to assign maliciousness scores to communication flows, relying on learned embeddings and structural patterns. The framework includes steps for node feature embedding, pattern learning, and classification. Additionally, they identify a potential data leakage issue in traditional evaluation methods and propose a solution to ensure robust validation [45]. Altaf et al. [46] proposed a botnet detection framework for IoT networks using a Sequential Gated Graph Convolutional Neural Network (GGCN). The model encodes network traffic as time-stamped multi-edge graphs, enabling it to capture temporal dependencies and hidden relationships. It employs a two-layer gated architecture with customised message-passing and aggregation functions optimised for time-series data [46]. He et al. [47] developed Illuminati; an explainable AI framework designed for cybersecurity applications using Graph Neural Networks. Without requiring prior knowledge of GNNs, Illuminati identifies key nodes, edges, and attributes influencing model predictions. It was evaluated on tasks like code and smart contract vulnerability detection, achieving 87.6% subgraph fidelity,a 10.3% improvement over existing methods. The explanations produced by Illuminati are also interpretable by domain experts, making it a valuable tool for trustworthy and transparent cybersecurity systems [47].

Additional research has explored the application of graph-based learning in broader cybersecurity and distributed environments. Narne [48] proposed a GNN-based methodology for IoT cybersecurity that models complex device interactions for real-time anomaly detection and threat mitigation, achieving high detection accuracy and reduced false positives [48]. Sareddy and Hemnath [49] introduced a federated learning framework integrating GNNs with Hashgraph technology, enabling privacy-preserving threat detection in distributed systems with low latency [49]. Other approaches have focused on advanced data representations for intrusion detection. Liu et al. [50] proposed Log2Vec, which transforms system logs into heterogeneous graph structures to capture complex relationships among events, improving the detection of insider threats and advanced persistent threats [50]. Similarly, Song et al. [51] developed DeepMem, a graph-based framework for analysing raw memory data, demonstrating the effectiveness of GNNs in identifying hidden structures within complex system environments [51]. Huo et al. [19] (2025) proposed a graph-based multi-agent reinforcement learning approach for autonomous air combat, combining GraphSAGE and optimization techniques to enable scalable and efficient decision-making [19]. Vyas et al.[52] (2025) surveyed autonomous cyber defence systems and identified key challenges such as scalability, explainability, and real-world deployment of intelligent security agents [52]. Ma et al. [53] (2024) proposed an adaptive graph neural network-based framework for dynamic task allocation in UAV–UGV systems, enabling efficient and flexible coordination in complex environments [53]. While these studies highlight the effectiveness of graph-based models across various cybersecurity domains, they primarily focus on detection, representation learning, or distributed security. In contrast, the proposed work emphasises the integration of graph-based intrusion detection with real-time UAV swarm coordination, addressing the need for unified cyber-physical defence mechanisms in dynamic operational environments.

# 3. Background

This section provides the theoretical foundation for the proposed framework by examining the role of Graph Neural Networks in modeling cyber-physical defense systems. It explores how GNNs can effectively represent complex relationships within dynamic environments such as drone swarms and battlefield communication networks. Key concepts, including inductive learning with GraphSAGE and graph-based modeling of UAV networks, are introduced to demonstrate the suitability of GNNs for intrusion detection, threat propagation analysis, and autonomous swarm coordination in multi-domain warfare scenarios.

## 3.1 Graph Neural Networks in Cyber-Physical Defense

Graph Neural Networks have emerged as powerful learning architectures for modeling and analyzing relational data. Unlike traditional deep learning models that rely on grid-like inputs (e.g., images or sequences), GNNs operate directly on graph-structured data, enabling them to capture both the topological relationships and feature-based dependencies between entities. This capability is particularly suited to cyber-physical systems, where interconnected agents such as drones, communication devices, and sensors interact dynamically over a network. In a battlefield or defense context, cyber intrusion detection systems (IDSs) face the challenge of recognizing sophisticated attack vectors—such as jamming, spoofing, or Advanced Persistent Threats (APTs)—which may not be discernible through isolated node behavior alone. By leveraging the inherent graph structure of communication networks, GNNs can identify anomalous behaviors through the patterns of information flow, temporal inconsistencies, and multi-hop dependencies among nodes. Let $G = (V, E)$ represent a cyber-physical graph, where $V$ is the set of nodes (e.g., UAVs, routers, base stations) and $E \subseteq V \times V$ is the set of edges indicating communication or command links. Each node $v_i \in V$ is associated with a feature vector $v_i \in \mathbb{R}^{\mathcal{F}}$, such as trust score, packet transmission rate, or latency. The objective is to learn a mapping:

$$f: (X, A) \longrightarrow Y \quad (1)$$

where $X \in \mathbb{R}^{N \times F}$ is the node feature matrix, $A \in \{0,1\}^{N \times N}$ is the binary adjacency matrix, and $Y \in \{0,1\}^{N}$ is the predicted label vector indicating whether each node is benign or compromised.

## 3.2 GraphSAGE: Inductive Learning for Dynamic Networks

While classical GNN models such as GCNs rely on full-graph training and do not generalize to unseen graphs, GraphSAGE offers an inductive learning framework capable of making predictions on previously unseen nodes or graph structures—essential for real-time battlefield applications where agents enter or exit the network. GraphSAGE, introduced by Hamilton et al. [54], was designed to enable inductive learning on unseen nodes and subgraphs. Unlike GCNs that rely on transductive message passing across the entire graph, as shown in the figure 1, GraphSAGE samples a fixed-size neighborhood and aggregates features from local structures to generate embeddings. The GraphSAGE update rule for a node $v_i$at layer $k$ is given by:

$$h_i^{(k)} = \sigma(W^{(K)} \cdot AGGREGATE^{(k)})(\{h_i^{(k-1)}\} \cup \{h_j^{(k-1)}, \forall\, j \in N(i)\})) \quad (2)$$

Where:

$h_i^{(k)}$ is the embedding of node $i$ at layer $k$.

$N(i)$ is the set of neighbors of node $i$.

$AGGREGATE^{(k)}$ is a differentiable aggregation function (e.g., mean, max, or LSTM-based).

$\sigma$ is non-linear activation (e.g., ReLU).

$W^{(K)}$ are learnable weights at layer $k$.

The initial node features are:

$$h_i^{(0)} = x_i \tag{3}$$

This process is repeated for $K$ layers, and final node representations $h_i^{(k)}$ are used for classification:

$$\hat{y}_i = \text{softmax}(W_{out} \cdot h_i^{(k)}) \tag{4}$$

This formulation enables the GNN to detect compromised nodes by learning how anomalies propagate through the structure over time.

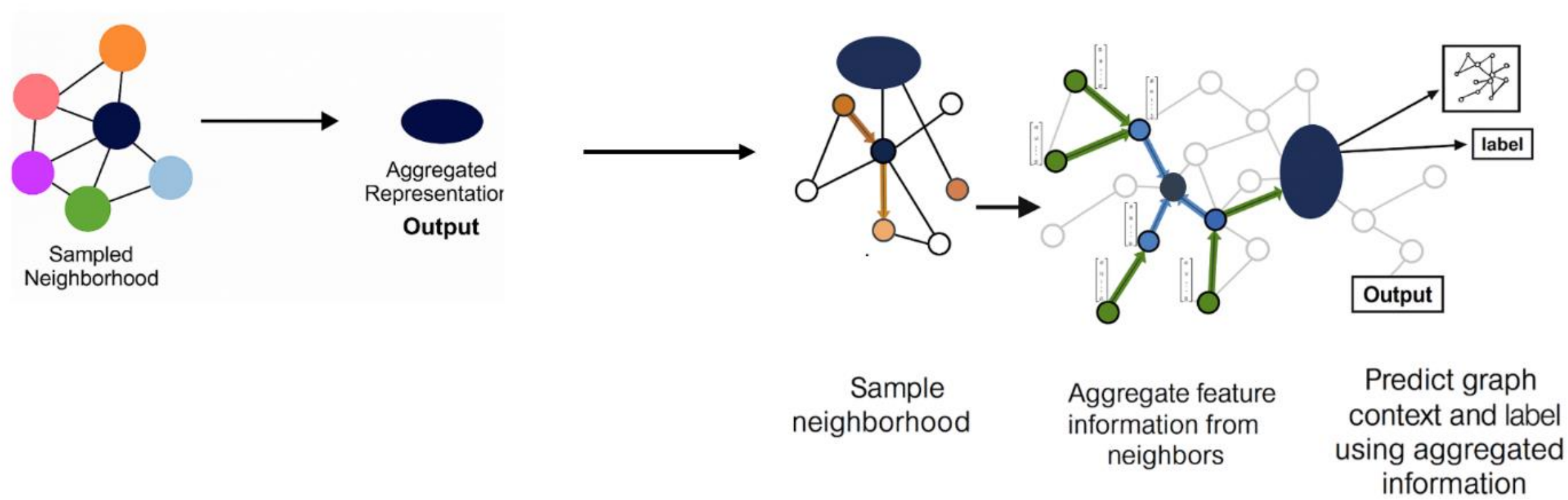

Figure 1. Visual illustration of the GraphSAGE sample and aggregate approach

## 3.3 UAV Swarm Modeling Graph

In response to cyber threats, Unmanned Aerial Vehicles (UAVs) must be rapidly re-tasked to respond, evade, or neutralize adversarial actions. This UAV network can be modeled as a dynamic swarm graph $G_s = (V_s, E_s)$, where:

1. Each node $v_i \in V_s$ is a drone equipped with sensors and communication modules.
2. Edges $(v_i, v_j) \in E_s$ represent communication or task dependencies.

Each drone node contains a time-varying feature vector:

$$x_i = [Trust, GPS\ Quality, Latency, Packet\ Rate, Battery, Orientation] \tag{5}$$

By applying GraphSAGE to the swarm graph, we can extract graph-level embeddings that inform:

1. Optimal drone tasking and interception planning.

2. Swarm reconfiguration around compromised nodes.
3. Autonomous evasion from jamming zones.

These decisions are driven by reinforcement-like policies that consume the high-dimensional embeddings learned by the GNN.

# 4. Suggested Method

In this section, we describe the GNN-based system architecture for intrusion detection and drone swarm coordination, first the system architecture and then the data used in training and testing.

## 4.1 Architectural Design Approach

The proposed method integrates a graph-based learning architecture with real-time cyber intrusion detection and drone swarm coordination for multi-domain defence applications. In this framework, a dynamic cyber-physical graph is constructed using simulated battlefield data representing nodes (connections, drones, or devices) and edges (data links, command channels, or communication dependencies). Each node incorporates temporal and behavioural properties such as packet activity, trust score, GPS signal integrity, and command latency. To detect threats, a GraphSAGE-based GNNis used to learn both structural and contextual information from the cyber-physical graph. The GNN model captures how anomalies propagate across interconnected entities and is trained to classify nodes as either benign or vulnerable (e.g., affected by jamming, spoofing, or advanced persistent threats). The trained model is then used in real time to detect malicious behaviour via incoming network traffic and drone telemetry. When malicious behaviour is detected, the framework autonomously initiates a drone-based countermeasure through the swarm coordination protocol. Drones are modelled as elements within a swarm graph and dynamically updated based on GNN predictions. The coordination module calculates optimal trajectories, target interception paths, and evasion manoeuvres using a reinforcement policy that leverages graph-level state embeddings from the GNN. This allows the swarm to autonomously neutralise threats, maintain formation integrity, and adapt to hostile environmental conditions.

To simulate and train this model, the system uses a complex drone cyber dataset, inspired by the Israel-Iran conflict, including loaded cyberattacks (such as advanced persistent attacks, distributed denial-of-service attacks, and GPS spoofing) and real-time drone telemetry. Raw communication logs are preprocessed and transformed into graph representations using NetworkX, with feature normalisation for each node and edge. Temporal attack patterns are encoded as time series snapshots, allowing the GNN to track evolving threats across iterations. The entire model is implemented using PyTorch Geometric, and simulations are run using a Python-based drone swarm environment. The framework is evaluated using four key metrics: intrusion detection accuracy, threat neutralisation rate, response latency, and swarm coordination. This integrated approach demonstrates the potential of graph neural networks (as a real-time decision engine, linking cyber intrusion detection to actual response in multi-domain conflict scenarios.

Figure 2 and Algorithm 1 offer the pseudocode and revised flowchart of the GNN-based intrusion detection and UAV swarm response approach, respectively, explaining the steps that lead to these results.

*Algorithm One: Demonstrates the formwork GNN-based intrusion detection and UAV swarm response approach pseudocode*

```
Initialize BattlespaceSimulation()
Initialize CyberPhysicalGraph()
Initialize DroneSwarm()
Load GNN_Model_Parameters()
Function Train_GNN_Model():
   For each simulation_scenario in SimulatedBattlespaceData:
      graph = Generate_Dynamic_CyberPhysicalGraph(simulation_scenario)
      node_features, edge_list = Extract_GraphFeatures(graph)
      labels = Label_ThreatNodes(graph)
      Train GNN_Model on (node_features, edge_list, labels)
   Evaluate GNN_Model on validation_data:
      Compute IntrusionDetectionAccuracy
      Compute ThreatNeutralizationRate
      Compute ResponseLatency
      Compute SwarmCoordinationEfficiency
   Save Trained GNN_Model
Function RealTime_Intrusion_Response():
   While system_active:
      new_data = Collect_RealTime_BattlespaceData()
      graph = Update_CyberPhysicalGraph(new_data)
      node_features, edge_list = Extract_GraphFeatures(graph)
      threat_predictions = GNN_Model.Predict(node_features, edge_list)
```

```
        If CompromiseDetected(threat_predictions):
            Trigger_DroneSwarm_Response(threat_predictions)
        Update_Evaluation_Metrics()
Function Trigger_DroneSwarm_Response(threat_predictions):
    For each compromised_node in threat_predictions:
        target_location = Get_Node_Location(compromised_node)
        Assign_Drone_To_Target(target_location)
    Compute_Optimal_Trajectories()
    Execute_Swarm_Coordination()
    Monitor_Mission_Success()
Function Update_Evaluation_Metrics():
    Log IntrusionDetectionAccuracy
    Log ThreatNeutralizationRate
    Log SwarmCoordinationPerformance
    Log SystemLatency
    If performance drops below threshold:
        Trigger Model_ReTraining()
If Evaluation_Metrics show degradation:
    Collect operational data
    Label data or use semi-supervised learning
    Retrain or fine-tune GNN model
    Redeploy updated model
```

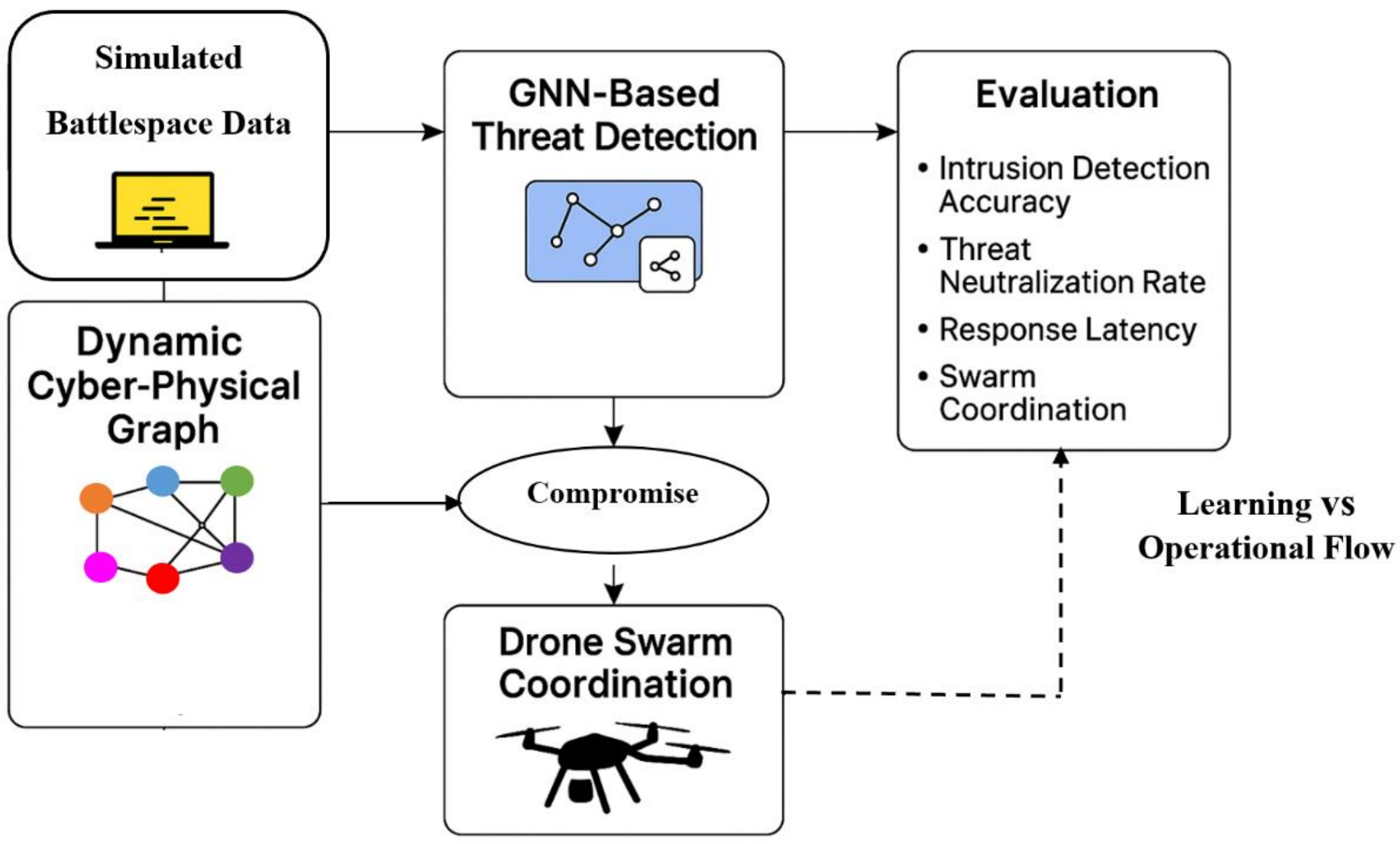

Figure 2. GNN-Based Intrusion Detection and UAV Swarm Response Architecture

## 4.2 Cyber drone dataset

The proposed framework uses a cyber-UAV dataset, specifically designed to simulate realistic battlefield conditions encountered in Israeli-Iranian cyber confrontations. The dataset is created through a controlled simulation process rather than being collected from a publicly available source. The dataset consists of two main components: structured telemetry and unstructured communications logs. The structured part includes features such as the UAV's GPS health, command response time, battery level, and swarm location, while the unstructured part consists of network messages, control signals, and system alerts extracted from the UAV's communication protocols. In the preprocessing phase, the unstructured text data is first encoded using natural language processing (NLP) techniques and embedded using domain-adapted vector encoders. Each token is converted to a numerical representation that reflects the average semantic frequency and contextual weight within the dataset. These combined features are aligned with their corresponding temporal and relational metadata to create a graphical representation of the cyber-physical battlefield. The resulting graph, where each node represents an entity (such as a drone, sensor, or controller) and each edge represents a communication link or dependency, is fed into a graph neural network (GNN) model, specifically GraphSAGE. This input enables the model to learn propagation patterns of advanced persistent threats (APTs), GPS spoofing, jamming attacks, and swarm-level intrusions.

By incorporating these threat features into a graph, the dataset becomes a powerful resource for training deep learning models that support intrusion detection and autonomous response for drones. The dataset effectively captures both cyberattack signatures and swarm behavioral anomalies, enabling real-time threat inference and drone coordination within the proposed GNN-based defense architecture. Figure 3 shows the Cyber UAV dataset screen.

cyber_drone_dataset_israeli_iranian_conflict_2025 - Notepad

File Edit Format View Help

| Drone_ID | Timestamp | GPS_Lat | GPS_Lon | Alt(m) | Bat(%) | Vel(m/s) | Latency(ms) | Loss(%) | Trust | Signal | Attack_Type | Threat |
|---|---|---|---|---|---|---|---|---|---|---|---|---|
| D001 | 2025-06-14T12:00:00Z | 34.747257 | 46.584933 | 136 | 95 | 15 | 149 | 0.18 | 0.65 | 0.66 | None | 0 |
| D002 | 2025-06-14T12:00:01Z | 34.744152 | 46.592381 | 139 | 87 | 15.3 | 94 | 0.2 | 0.81 | 0.58 | Spoofing | 1 |
| D003 | 2025-06-14T12:00:02Z | 34.744232 | 46.587425 | 153 | 86 | 14.2 | 88 | 0.03 | 0.89 | 0.54 | Jamming | 1 |
| D004 | 2025-06-14T12:00:03Z | 34.741436 | 46.593863 | 143 | 79 | 13.6 | 94 | 0.01 | 0.45 | 0.91 | Spoofing | 1 |
| D005 | 2025-06-14T12:00:04Z | 34.742433 | 46.592577 | 145 | 90 | 15.1 | 90 | 0.11 | 0.71 | 0.49 | APT | 1 |
| D006 | 2025-06-14T12:00:05Z | 34.748577 | 46.582213 | 148 | 72 | 14.3 | 88 | 0.05 | 0.47 | 0.73 | APT | 1 |
| D007 | 2025-06-14T12:00:06Z | 34.749828 | 46.595887 | 156 | 87 | 12.8 | 96 | 0.18 | 0.82 | 0.64 | DDoS | 1 |
| D008 | 2025-06-14T12:00:07Z | 34.750886 | 46.595097 | 141 | 82 | 12.9 | 112 | 0.09 | 0.62 | 0.41 | APT | 1 |
| D009 | 2025-06-14T12:00:08Z | 34.755459 | 46.584073 | 131 | 70 | 14.9 | 106 | 0.16 | 0.69 | 0.56 | Spoofing | 1 |
| D010 | 2025-06-14T12:00:09Z | 34.755477 | 46.597155 | 134 | 98 | 12.6 | 147 | 0.26 | 0.6 | 0.74 | Jamming | 1 |
| D011 | 2025-06-14T12:00:10Z | 34.749717 | 46.583008 | 146 | 91 | 14.9 | 88 | 0.25 | 0.7 | 0.78 | APT | 1 |
| D012 | 2025-06-14T12:00:11Z | 34.746367 | 46.584164 | 144 | 72 | 13.3 | 102 | 0.02 | 0.38 | 0.53 | DDoS | 1 |
| D013 | 2025-06-14T12:00:12Z | 34.742538 | 46.584974 | 159 | 72 | 13.8 | 81 | 0.23 | 0.39 | 0.44 | DDoS | 1 |
| D014 | 2025-06-14T12:00:13Z | 34.740149 | 46.588375 | 134 | 99 | 14.2 | 89 | 0.15 | 0.5 | 0.69 | APT | 1 |
| D015 | 2025-06-14T12:00:14Z | 34.743219 | 46.581554 | 144 | 89 | 15.5 | 122 | 0.27 | 0.42 | 0.93 | DDoS | 1 |
| D016 | 2025-06-14T12:00:15Z | 34.750289 | 46.59039 | 135 | 72 | 12.2 | 138 | 0.26 | 0.83 | 0.43 | DDoS | 1 |
| D017 | 2025-06-14T12:00:16Z | 34.758529 | 46.592589 | 151 | 73 | 14.8 | 146 | 0.07 | 0.75 | 0.82 | None | 0 |
| D018 | 2025-06-14T12:00:17Z | 34.745688 | 46.59378 | 160 | 78 | 13.2 | 110 | 0.29 | 0.68 | 0.91 | APT | 1 |
| D019 | 2025-06-14T12:00:18Z | 34.742946 | 46.588011 | 136 | 71 | 13 | 123 | 0.03 | 0.56 | 0.7 | None | 0 |
| D020 | 2025-06-14T12:00:19Z | 34.746089 | 46.587768 | 137 | 98 | 14.4 | 82 | 0.13 | 0.68 | 0.52 | DDoS | 1 |
| D021 | 2025-06-14T12:00:20Z | 34.743427 | 46.58089 | 144 | 88 | 15.5 | 86 | 0.03 | 0.77 | 0.71 | None | 0 |
| D022 | 2025-06-14T12:00:21Z | 34.758875 | 46.588517 | 140 | 100 | 13.4 | 116 | 0.25 | 0.69 | 0.92 | DDoS | 1 |
| D023 | 2025-06-14T12:00:22Z | 34.754304 | 46.596322 | 136 | 80 | 13.4 | 114 | 0.15 | 0.91 | 0.62 | DDoS | 1 |
| D024 | 2025-06-14T12:00:23Z | 34.741985 | 46.581329 | 149 | 76 | 14.6 | 120 | 0.03 | 0.71 | 0.55 | Spoofing | 1 |
| D025 | 2025-06-14T12:00:24Z | 34.749364 | 46.597858 | 130 | 83 | 15.6 | 112 | 0.18 | 0.98 | 0.56 | Spoofing | 1 |
| D026 | 2025-06-14T12:00:25Z | 34.752797 | 46.595884 | 133 | 70 | 13.8 | 87 | 0.28 | 0.98 | 0.73 | None | 0 |
| D027 | 2025-06-14T12:00:26Z | 34.758711 | 46.58523 | 140 | 86 | 13.8 | 124 | 0.14 | 0.37 | 0.74 | Jamming | 1 |
| D028 | 2025-06-14T12:00:27Z | 34.753644 | 46.581308 | 140 | 92 | 14.5 | 136 | 0.12 | 0.63 | 0.48 | None | 0 |
| D029 | 2025-06-14T12:00:28Z | 34.751467 | 46.592587 | 151 | 73 | 13 | 150 | 0.29 | 0.76 | 0.81 | Jamming | 1 |
| D030 | 2025-06-14T12:00:29Z | 34.753506 | 46.583605 | 156 | 76 | 12.8 | 81 | 0.28 | 0.83 | 0.78 | DDoS | 1 |
| D031 | 2025-06-14T12:00:30Z | 34.758235 | 46.584079 | 142 | 95 | 12 | 111 | 0.06 | 0.83 | 0.54 | None | 0 |
| D032 | 2025-06-14T12:00:31Z | 34.741908 | 46.595477 | 156 | 96 | 16 | 135 | 0.16 | 0.62 | 0.83 | APT | 1 |
| D033 | 2025-06-14T12:00:32Z | 34.757994 | 46.580088 | 130 | 82 | 13.8 | 110 | 0.13 | 0.79 | 0.41 | APT | 1 |
| D034 | 2025-06-14T12:00:33Z | 34.757383 | 46.593512 | 143 | 92 | 14.9 | 137 | 0.1 | 0.6 | 0.85 | Jamming | 1 |
| D035 | 2025-06-14T12:00:34Z | 34.740393 | 46.590289 | 135 | 71 | 14 | 102 | 0.04 | 0.59 | 0.91 | APT | 1 |
| D036 | 2025-06-14T12:00:35Z | 34.748918 | 46.590402 | 152 | 71 | 12.4 | 101 | 0.25 | 0.39 | 0.81 | Spoofing | 1 |
| D037 | 2025-06-14T12:00:36Z | 34.7587 | 46.59491 | 143 | 94 | 14.5 | 97 | 0.24 | 0.35 | 0.72 | Jamming | 1 |
| D038 | 2025-06-14T12:00:37Z | 34.754318 | 46.579171 | 150 | 75 | 13.7 | 128 | 0.02 | 0.86 | 1 | None | 0 |
| D039 | 2025-06-14T12:00:38Z | 34.758982 | 46.579063 | 136 | 87 | 14.1 | 83 | 0.27 | 0.93 | 0.51 | APT | 1 |
| D040 | 2025-06-14T12:00:39Z | 34.752397 | 46.581487 | 147 | 74 | 14.6 | 115 | 0.25 | 0.36 | 0.44 | None | 0 |
| D041 | 2025-06-14T12:00:40Z | 34.747081 | 46.57997 | 156 | 99 | 12.4 | 137 | 0.06 | 0.99 | 0.78 | Spoofing | 1 |
| D042 | 2025-06-14T12:00:41Z | 34.752772 | 46.594223 | 157 | 78 | 12.4 | 82 | 0.15 | 0.93 | 0.45 | Spoofing | 1 |

Figure 3. Cyber drone dataset screen

## 4.3 Cyber-Drone Dataset Description

The experiments were conducted on a simulated electronic drone dataset designed to represent drone telemetry data and physical cyber threat scenarios. The dataset was designed to reflect real-world communication patterns and attack behaviors commonly observed in drone cybersecurity studies. It is inspired by documented drone operation incidents and cyberattacks, including GPS spoofing, jamming, and coordinated cyber intrusions reported between June 13 and 24, 2025, while maintaining a controlled simulation environment and not a general reference dataset [55][56][57].

Each drone status record is time-stamped and includes key physical cyber parameters required for intrusion detection and threat classification. The dataset integrates structured telemetry data (such as GPS coordinates, speed, battery level, latency, packet loss, confidence score, and signal strength) with communication-related information, such as control messages and system alerts. Table 1 shows the cyber drone telemetry dataset fields and descriptions.

**Table 1. Cyber drone telemetry dataset fields and descriptions**

| Field | Description |
|---|---|
| Drone_ID | Unique identifier for each UAV |
| Timestamp | ISO 8601 time format indicating when the data was recorded |
| GPS_Lat | Latitude coordinate of the UAV |
| GPS_Lon | Longitude coordinate of the UAV |
| Alt(m) | Altitude in meters |
| Bat(%) | Battery level percentage |
| Vel(m/s) | UAV velocity in meters per second |
| Latency(ms) | Communication latency in milliseconds |
| Loss (%) | Packet loss percentage |
| Trust | Trust score assigned to the UAV's communication behavior |
| Signal | Signal strength reading |
| Attack_Type | Type of detected cyberattack (e.g., Spoofing, Jamming, None) |
| Threat | Binary label (1 = compromised/threat detected, 0 = normal) |

#### 4.3.1. Dataset Scale

The dataset contains approximately 8,000 samples, each representing the state of a drone at a specific point in time. Each sample includes characteristics such as GPS location, response time, signal strength, packet loss, confidence level, and battery level.

#### 4.3.2. Category Classifications

Each sample is assigned to one of four classes:

1. Benign
2. Jamming
3. Spoofing
4. Advanced Persistent Threat (APT)

The classifications are created based on predefined simulation rules that correspond to attack conditions (e.g., abnormal response time for jamming, GPS deviation for identity spoofing).

#### 4.3.3. Data Segmentation

The dataset is divided into:

1. Training set: 70%

2. Validation set: 10%
3. Test set: 20%

All models are trained and evaluated using the same segmentation to ensure a fair comparison.

#### 4.3.4. Category Distribution

The data is distributed almost evenly across the four categories. Minor variations may occur due to the randomness of the simulation.

#### 4.3.5. Graph Construction

The dataset is transformed into a graph representation $G = (V, E)$, where:

1. Each node represents a drone.
2. The node's characteristics correspond to its telemetry attributes.

#### 4.3.6. Edge Generation

Edges are created based on:

1. Spatial proximity (threshold distance between drones).
2. Similarity in communication response time.

Two nodes are connected if:

1. Distance < a predefined threshold (d).
2. Response time difference < a threshold (l).

#### 4.3.7. Time-based modeling

The dataset is processed in discrete time phases. The graphs are updated at each time phase to reflect changes in node and connection states.

#### 4.3.8. Data Classification

Classifications are added during simulation based on the input attack scenarios:

1. Jamming: Associated with increased latency and signal degradation.
2. Spoofing: Associated with abnormal GPS deviations.
3. Advanced Persistent Attack (APT): Associated with a shared vulnerability across multiple attributes.

#### 4.3.9. Simulation Parameters

The simulation includes the following:

1. A variable number of drone nodes per time step
2. Random attack injection
3. Telemetry data distortion to simulate real-world variability

All models (GraphSAGE, GCN, GAT) are trained under identical conditions using the same dataset, graph construction process, and evaluation protocol.

#### 4.3.10. Reproducibility and Data Availability

The dataset used in this study is simulation-based and was generated using predefined rules to model the physical behavior of cyber drones under various attack scenarios. The generation process includes telemetry simulation, attack injection (jamming, identity spoofing, and advanced persistent threat (APT) patterns), and graph construction based on spatial and communicative relationships. To ensure reproducibility, all experimental parameters, including feature definitions, graph construction rules, neighborhood sampling, and training settings, are explicitly described in this study. Although the dataset is not derived from a general reference standard, it can be reproduced using the described simulation procedure. Future work will focus on publishing a portion of the dataset or data generation scripts to support further research and validation.

# 5. Methodology

This section outlines the complete methodology used to develop, train, and evaluate the proposed GNN-based cyber-physical defense system. It details the process of constructing the cyber-physical graph, loading and preprocessing the dataset, extracting features, designing the graph neural network model, coordinating the drone swarm, and assessing system performance through both metrics and visualisation tools.

## 5.1 Cyber-Physical Graph Construction

The framework begins by simulating and constructing a cyber-physical graph that captures battlefield dynamics. To accurately represent the complex battlefield dynamics of modern cyber-physical warfare, the uploaded cyber drone dataset undergoes a careful transformation into a cyber-physical graph. This process involves identifying key entities, such as drones, command centres, communication relays, and sensors, which are modelled as nodes. Interconnections, such as communication links, command dependencies, and trust relationships, are modelled as edges in the graph. Figure 4 shows the Cyber-Physical Graph. Using the NetworkX Python library, these raw telemetry logs are processed and linked to graph structures that encode topological and contextual information. Each node is enriched with temporal and behavioural properties, including timestamped GPS location, battery level, speed, latency, trust score, signal integrity, and any associated cyberattack classifications (e.g., spoofing, jamming, and advanced persistent threats). These properties are essential for detecting and visualising intrusion patterns and for enabling subsequent graph neural network models to learn threat propagation. To improve interpretability and visualisation, Matplotlib has been integrated to display the resulting cyber-physical graph. This graphical representation enables researchers and defence analysts to visually track threat propagation, identify vulnerabilities, and monitor the structural integrity of the battlefield communications network. Ultimately, the generated cyber-physical graph serves as the primary input for GNN-based threat detection and drone swarm coordination.

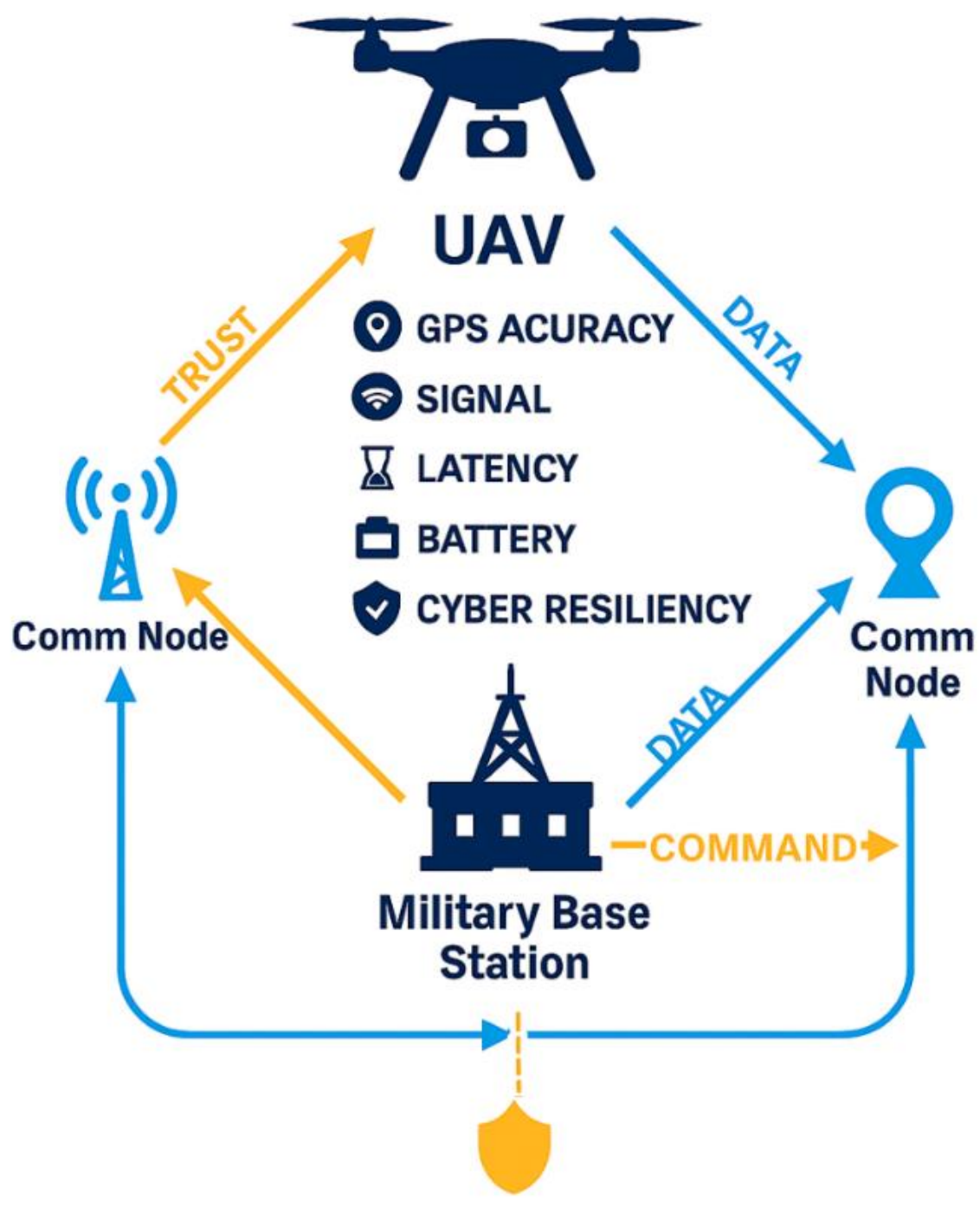

Figure 4. Cyber-Physical Graph

## 5.2 Data loading

To support cyber-physical threat analysis in simulated conflict environments, the system enables seamless import and management of telemetry data through an integrated interface. Users can upload a drone dataset, which includes historical drone telemetry, GPS coordinates, latency, packet loss, confidence scores, and attack classifications (e.g., spoofing, jamming, and advanced persistent threats). When selecting a dataset, the framework uses the Pandas library to efficiently parse and load the data into structured data frames. This process ensures compatibility with graph-based learning models. To improve usability, the system displays a snapshot of the raw dataset rows, enabling users to verify the integrity and structure of the data before preprocessing. Each record is then preprocessed to normalise numerical attributes and code categorical threat types. This step ensures consistency in feature measurement and prepares the data for transformation into a cyber-physical graph structure, where drones and infrastructure components are treated as nodes, and their communication and trust dependencies are treated as edges. The resulting structural representation becomes the basis for intrusion detection and swarm coordination based on graph neural networks. Figure 5 shows the data loading and preprocessing pipeline for the cyber-drone telemetry dataset.

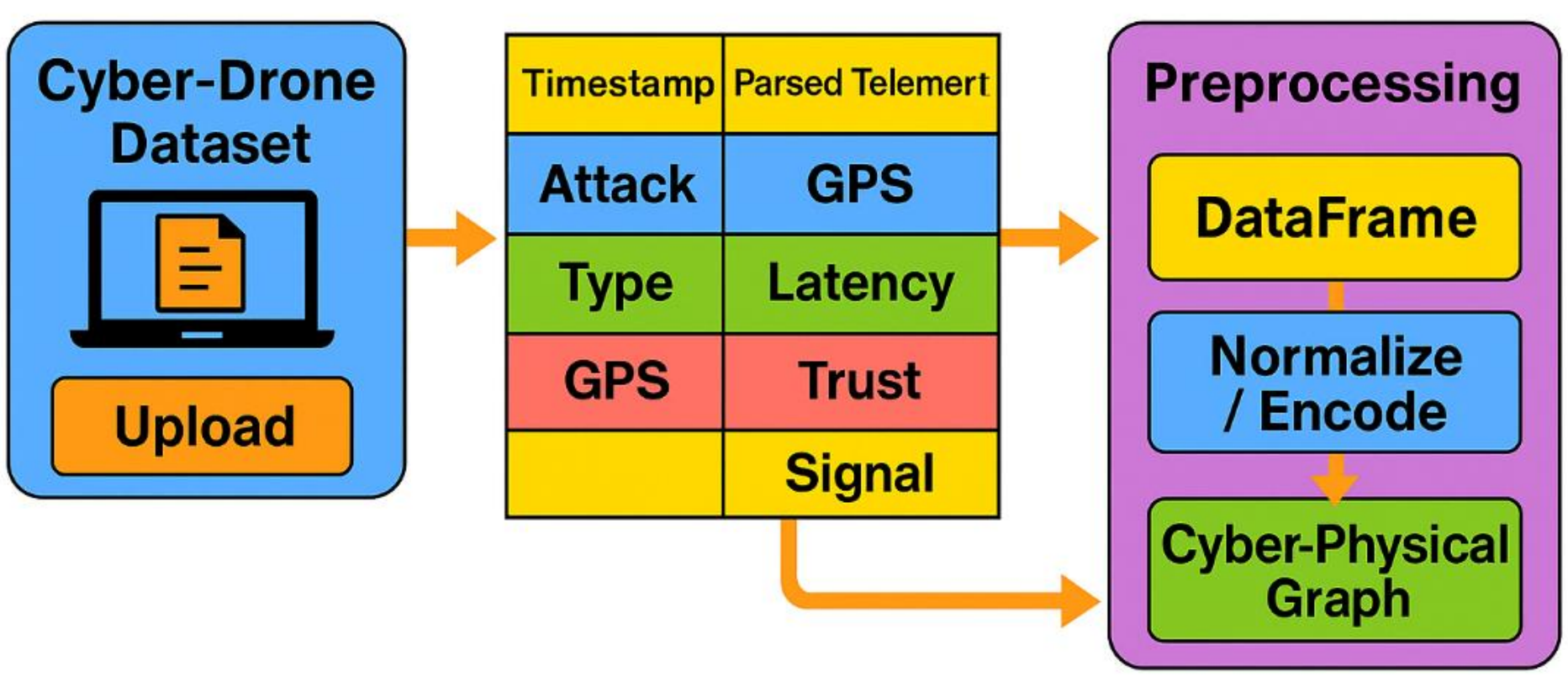

Figure 5. Data loading and preprocessing pipeline for the cyber-drone telemetry dataset

## 5.3 Dataset Preprocessing

Before creating a graph neural network (GNN) model using the drone telemetry dataset an extensive data preprocessing phase must be conducted to enable quality and consistency of data upon which the user will perform further graphical analysis on cyber drones. In addition, this data preprocessing phase will assist with addressing any quality issues associated with the data; this will be beneficial to the user as he or she will have access to a new revenue-generating capability from working in an unfamiliar industrial (i.e. cyber-physical) environment via the development of an overall aggregated dataset that can also serve as a conduit for new identifier-feature data which will be used for modelling purposes. Each one of the different data attributes contained within the drone telemetry datasets (including but not limited to GPS location coordinates, remaining battery power levels, latency, and signal strength) may have a null (missing) or blank (empty) value associated with them. Starting now, every individual electrical data attribute that has a 'null' or 'blank' value will have that value converted into some usable number through one of several methods (e.g., substituting "0" for the missing value, statistical inference, etc.) so as not to jeopardize successful feature extraction. Additionally, categorical values - such as data elements representing types of attacks (for instance, spoof, jam, or advanced persistent threat) or types of binary threat - will be converted into numeric values using Scikit-learn's LabelEncoder feature so they can be used with PyTorch model definitions created from telemetry data sets. After the standardised template is developed, all the telemetry data collected from various nodes will be divided into two groups, the telemetry data plus information about how a node behaves (X: also referred to as input data) and the data indicating whether a network node was attacked or not (Y: also referred to as output). In this section, I will explain how I built and trained a graph neural network to determine whether a node has been attacked based on its behaviour relative to the behaviour of the nodes in the dataset before and after an attack. The figure below shows how the data processing pipeline was created to process the telemetry data prior to the implementation of the Graph Neural Networks. See Figure 6 below for a flow chart of how this data processing pipeline looked like in terms of the data collection, processing and storage operations. Coordinating drones in real time during the Iran-Israel conflict viasimulation of pre-processing would provide precise detection of actual threats and be essential for military users, emergency management agencies involved, local governments and municipalities.

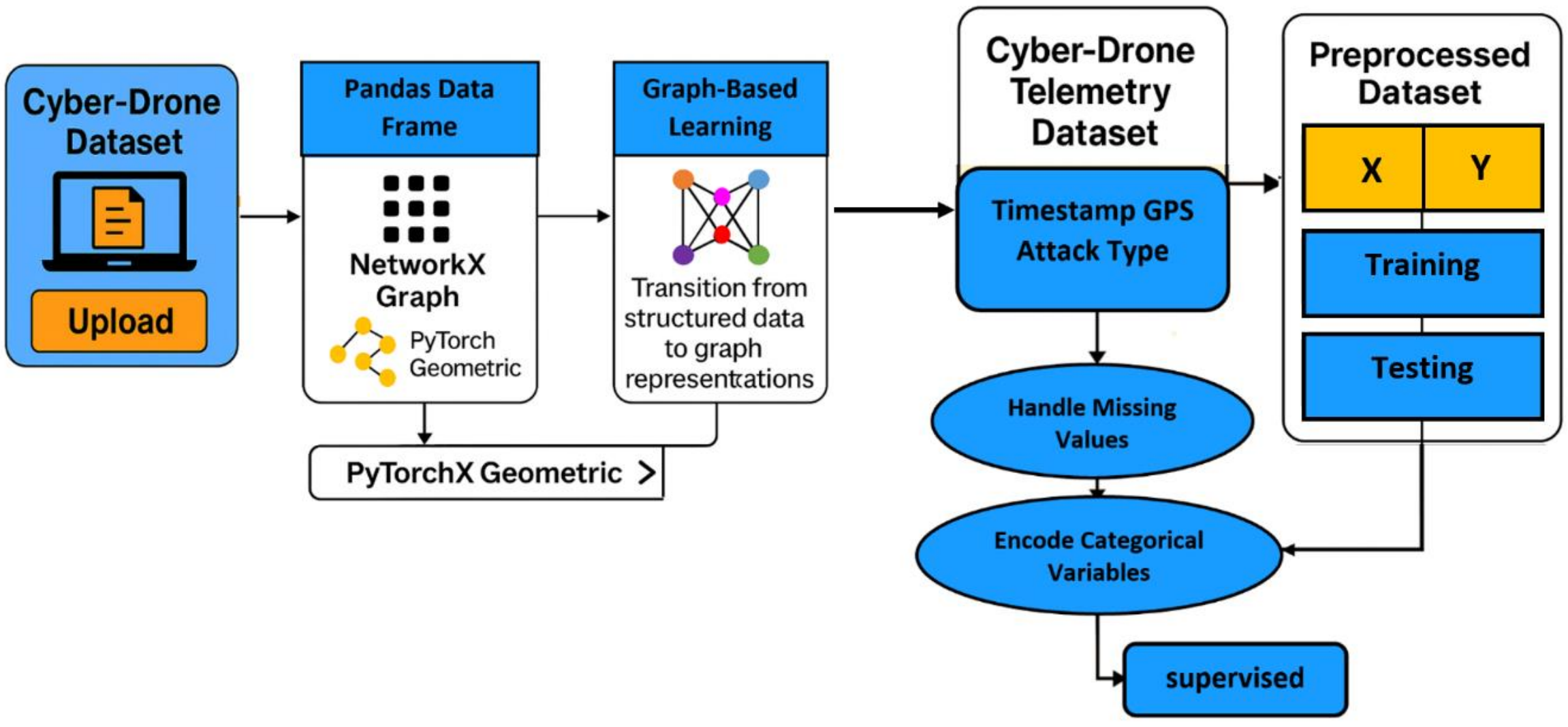

Figure 6. Data processing pipeline

## 5.4 Feature Extraction and Graph Representation

To effectively capture the cyber-physical dynamics within the simulated Israeli-Iranian conflict environment, we employed Python’s NetworkX library to transform raw telemetry logs and communication metadata into structured graph representations suitable for downstream graph neural network (GNN) processing. Each UAV (Drone_ID) is represented as a node in the graph, encapsulating a rich set of features derived from system-level data streams. These include geospatial coordinates (GPS_Lat, GPS_Lon, Altitude), performance indicators (Velocity, Battery Level), network metrics (Signal Strength, Communication Latency, Packet Loss), and security-related features (Trust Score, Threat Label). These multidimensional attributes allow the graph to holistically reflect each drone's physical state, network behaviour, and trustworthiness in real time.

Edges in the graph encode inter-UAV relationships and are dynamically established based on proximity thresholds, latency correlations, or command-and-control linkages. Edge features, such as latency and packet loss, provide contextual understanding of the communication reliability and serve as valuable indicators for detecting compromised links or nodes. To ensure training stability and cross-scenario generalisation, both node and edge features undergo min-max normalisation and are aligned to a unified scale. This preprocessing step guarantees compatibility with the GNN model and preserves the semantic relevance of each input dimension. Moreover, the graph structure is continuously updated across temporal snapshots, capturing the evolution of threat landscapes over time. This dynamic graph construction enables the system to model spatio-temporal dependencies, which are critical for recognising attack propagation patterns such as GPS spoofing, signal jamming, or coordinated drone compromise. Figure 7 shows the flowchart of feature extraction and graphical representation The resulting graph-structured data not only reflects the topological and behavioural complexity of the cyber-drone environment but also serves as a foundational input for threat detection models such as GraphSAGE, GCN, or GAT, enabling interpretable and mission-adaptive cyber defence capabilities.

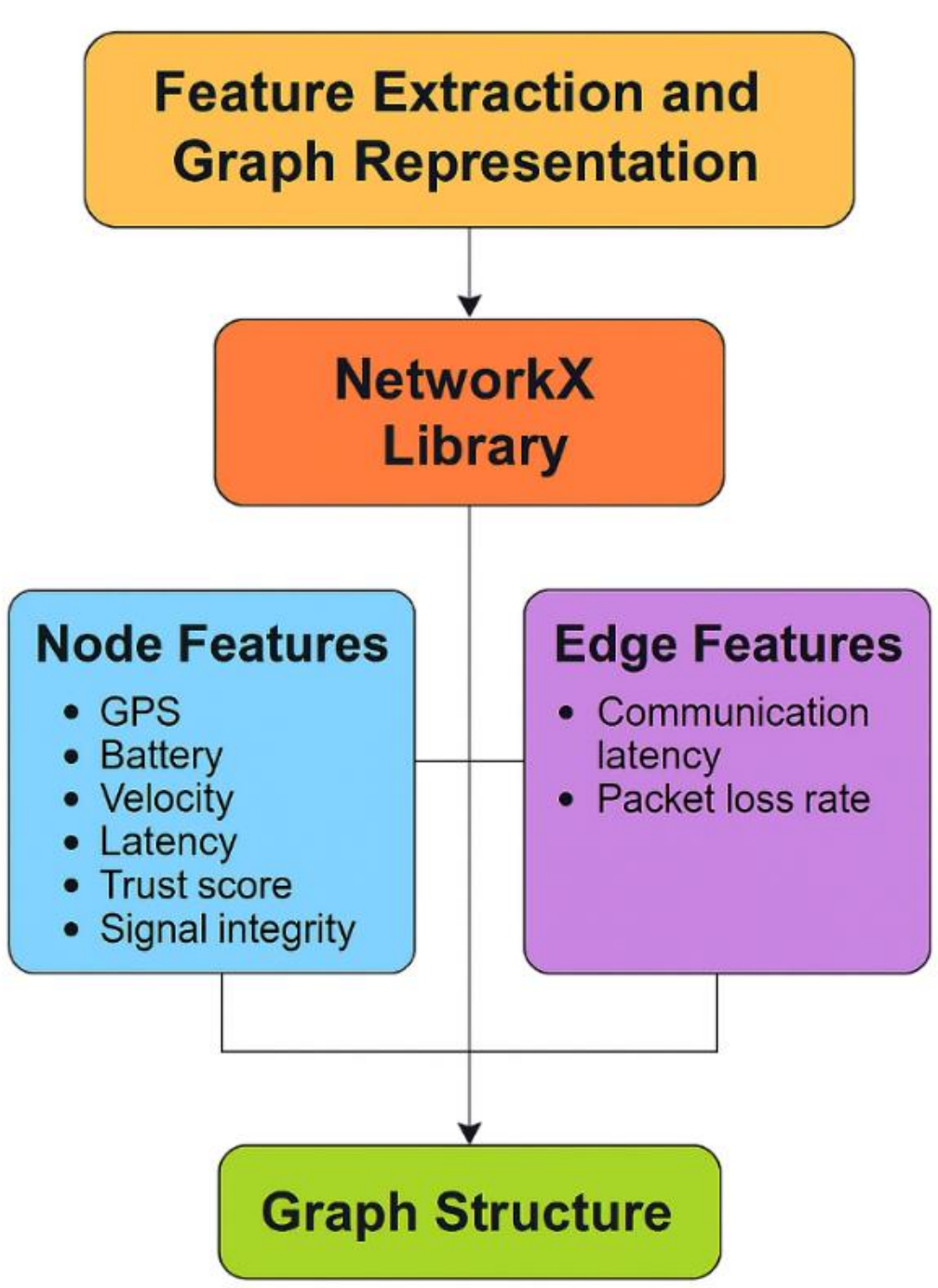

Figure 7. Flowchart of feature extraction and graphical representation

## 5.5 Graph Neural Network-Based Threat Detection

To detect and respond to complex cyber threats in real-time battlefield scenarios, the framework implements a graph neural network-based intrusion detection system specifically designed to analyse cyber-physical interactions within a drone network. The selected GNN model, GraphSAGE, is designed to support inductive learning on realistic dynamic graphs, enabling scalable threat inference even as new aircraft and devices are introduced during operations. The GNN architecture begins with an input layer that ingests node properties derived from a pre-processed cyber drone dataset, such as GPS location, battery level, confidence score, latency, signal strength, and attack indicators. These properties are passed through multiple neighbour aggregation layers, where each node's representation is updated by sampling and aggregating information from its neighbours using a mean-based aggregation function, ensuring computational efficiency and stability. To enhance the model's robustness and generalization, dropout layers are strategically applied between GNN layers, reducing the risk of overfitting. The final output layer applies a Softmax activation function to classify each node (drone or device) as benign or compromised, depending on the presence and nature of cyberattacks (e.g., spoofing, jamming, advanced persistent threats).

The model is compiled using an Adam optimiser and a class cross-entropy loss function, reflecting the multi-class nature of threat classification. Training is performed over multiple periods using minibatch stochastic gradient descent. The training phase leverages both node characteristics and the structural graph context, while validation is performed on an excluded subset to monitor generalisation performance. The model is designed with computational efficiency in mind, enabling potential deployment on resource-constrained edge devices through sampling-based graph processing. Figure 8 shows the graph neural network-based threat detection To ensure optimal detection accuracy, model checkpoints are used to retain the best-performing GNN configuration during training. The trained GNN model is then integrated into the drone swarm coordination module, enabling real-time anomaly detection and adaptive mission planning in response to emerging threats.

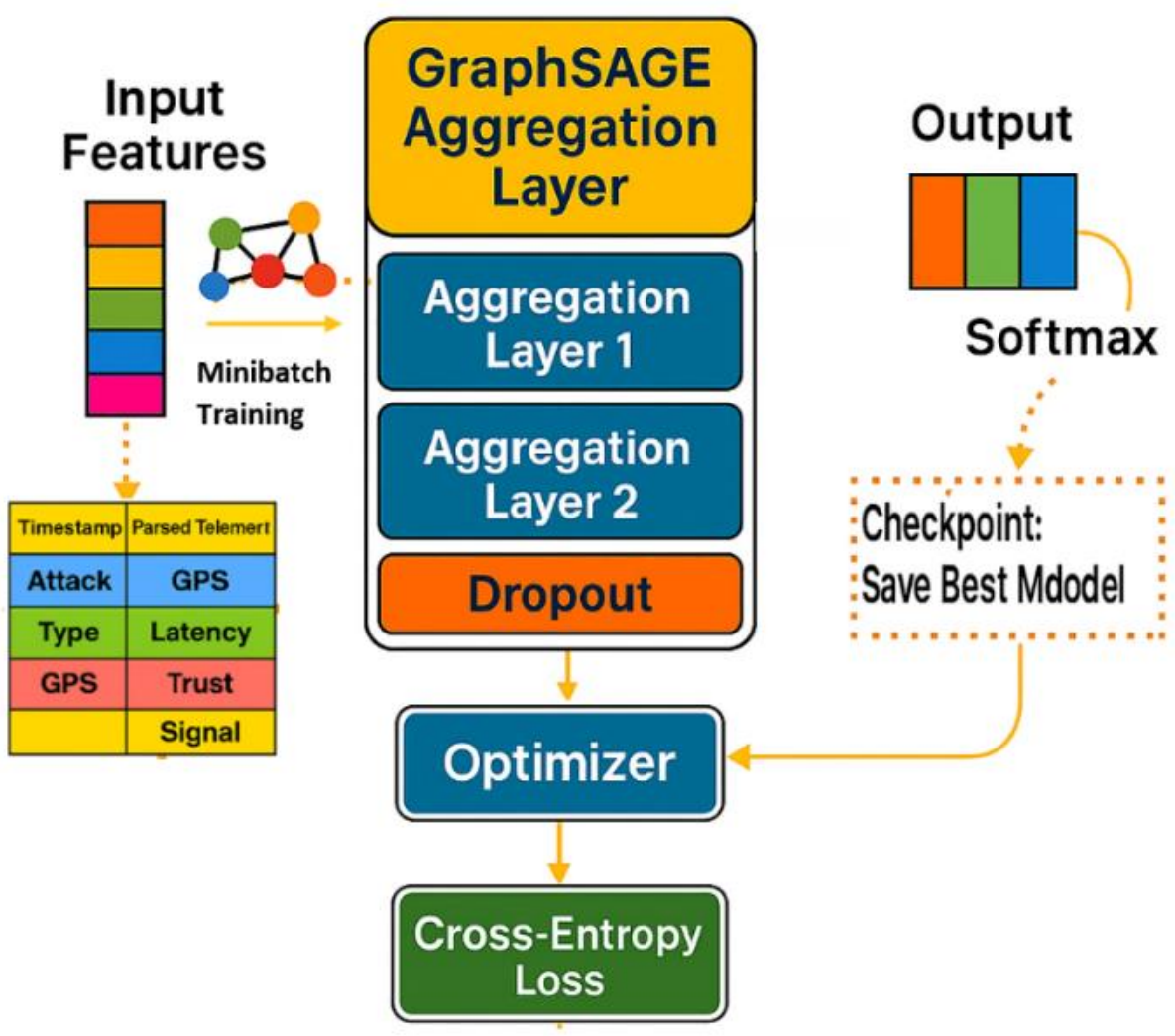

Figure 8. Graph neural network-based threat detection

### 5.5.1. Model implementation

To ensure reproducibility, the implementation details of the GraphSAGE-based model are provided as follows: -

#### 1. GraphSAGE Structure

The model consists of two GraphSAGE layers followed by a fully connected classification layer. Each GraphSAGE layer uses an average sum to combine neighborhood information.

1. Number of layers: 2
2. Hidden dimensions: 64
3. Output dimensions: 4 (corresponding to the four classes)
4. Activation function: ReLU

#### 2. Neighbor Sampling

For each node, a fixed number of neighbors is chosen to ensure computation efficiency and learning stability:

1. Layer 1: 15 neighbors
2. Layer 2: 10 neighbors

These values were chosen to balance information gathering and computational cost, while avoiding excessive smoothing in deeper layers.

#### 3. Input Characteristics

Each node represents an F-dimensional characteristic vector, which includes the following:

1. GPS coordinates (2)
2. Speed (1)
3. Battery level (1)
4. Response time (1)
5. Packet loss (1)
6. Confidence score (1)
7. Signal strength (1)

Total feature dimensions: 8

All features are calibrated using minimum and maximum normalization to the range [0,1].

## 4. Training Setup

The model was trained using the following settings:

1. Loss Function: Cross-entropy Loss
2. Optimizer: Adam
3. Learning Rate: 0.001
4. Batch Size: 64
5. Number of Training Cycles: 100

Early stop is applied based on validation loss.

## 5. Classification Output

The final layer applies a softmax function to produce classification probabilities for the following:

1. Non-harmful
2. Spoofing
3. Impersonation
4. Advanced persistent threat

## 6. Drone Response Map

The output of a graphical neural network is used to guide drone movements through a rules-based coordination mechanism:

1. No Action Required: No action is required.
2. Jambling: The communication path is altered.
3. Spoofing: Navigation parameters are modified.
4. Advanced Persistent Threats: The affected node is isolated, and a defensive response is activated.

This map enables the system to translate node-level predictions into system-level actions.

## 7. Implementation Framework

The model was implemented using PyTorch Geometric, and all experiments were conducted under identical settings for the GraphSAGE, GCN, and GAT models.

## 5.6 Drone Swarm Coordination and Response System

The coordination mechanism employs a rules-based methodology rather than reinforcement learning to ensure interpretability and repeatability. After a threat is successfully identified by a graphical neural network-based intrusion detection module, the system transitions to active defense by invoking the drone swarm coordination engine. This module interprets categorized node states (e.g., benign, jamming, spoofing, or compromised) and translates them into actionable strategies for autonomous drones operating within a networked swarm formation. Each drone within the swarm is modeled as a node in a dynamic swarm graph, where edges represent communication links, spatial proximity, or shared mission objectives. This graph is continuously updated to reflect changing conditions, node failures, or newly detected anomalies. Spatial and contextual implications generated by the GraphSAGE model are incorporated into the swarm graph as node characteristics, encoding information related to the threat and the environmental context. The outputs of the GraphSAGE model are used to guide drone operations through a rules-based coordination mechanism. Each node in the graph is categorized into one of four categories: benign, jamming, spoofing, or advanced persistent threat. Based on the predicted category, predefined response strategies are applied:

1. Benign: No action is required, and operations continue normally.
2. Jamming: Communication paths are rerouted to maintain connectivity.
3. Spoofing: Navigation parameters are modified to mitigate GPS inconsistencies.
4. Advanced persistent threat: The affected node is isolated, and a defensive response is activated.

This mechanism enables the system to translate node-level threat predictions into system-level drone coordination actions. The coordination mechanism is deterministic and does not rely on reinforcement learning, ensuring transparency and repeatability.

## 5.7 Performance Evaluation

After training in a graph neural network (GNN)-based threat detection model, its operational effectiveness is systematically evaluated using a comprehensive set of performance metrics. Key indicators, such as accuracy, precision, recall, and F1 score, are calculated to measure the model's classification performance in detecting various forms of cyber threats, including spoofing, obfuscation, and advanced persistent threats (APTs), in drone swarm environments. To improve interpretability, confusion matrices are constructed, visually displaying the distribution of correct and incorrect classifications across threat categories. These matrices provide practical insights into the model's robustness and error patterns, which aid in iterative model improvement. In addition to classification accuracy, the system also measures response time and threat neutralisation rate, which are vital for real-time military operations. These metrics reflect how quickly and effectively the GNN can identify cyber-physical vulnerabilities within the drone network and activate responses. Furthermore, the trained GNN model is applied to unprecedented test data from the cyber drone dataset, allowing for a realistic assessment of its generalizability in dynamic conflict scenarios. This process ensures not only that the model performs well in training but also that it maintains its reliability and adaptability under simulated Israeli-Iranian cyber warfare conditions.

### 5.7.1. Simulation Environment

Drone swarm coordination is evaluated in a simulation environment that models basic communication and interaction behaviors among UAVs. Each UAV is represented as a node in a dynamic graph, with edges reflecting communication links and spatial proximity relationships. The simulation includes:

1. Simplified spatial UAV movement in a two-dimensional environment.
2. Communication constraints based on latency and signal strength.
3. Dynamic updates of node states based on detected threats.
4. No explicit physical dynamics or collision modeling, as the focus is on communication-level coordination.

This setup enables controlled evaluation of coordination strategies while maintaining computational efficiency.

### 5.7.2. Definition of Runtime Metrics

To ensure clarity and reproducibility, the system performance metrics are defined as follows:

1. Accuracy, Precision, Recall, and F1-score: computed using standard classification formulas based on true positives, false positives, true negatives, and false negatives.
2. Threat Neutralisation Rate: defined as the percentage of detected threats for which a corresponding response action is successfully triggered by the coordination mechanism.
3. Response Time: defined as the average time interval between threat detection by the graph neural network (GNN) model and the initiation of the corresponding UAV response.
4. Swarm Coordination Effectiveness: measured as the percentage of scenarios in which UAVs successfully adapt their configuration (e.g., rerouting or node isolation) without communication failure.

All metrics are computed on the test dataset using consistent evaluation conditions across all models.

### 5.8 Visualisation

To provide an in-depth understanding of the effectiveness and reliability of the graph neural network (GNN)-based intrusion detection and drone coordination framework, a series of comparative visualisations were created. These visualisations play a crucial role in analysing the performance of the proposed model across multiple dimensions relevant to real-time battlefield scenarios. Bar charts and performance graphs were created using Matplotlib, illustrating evaluation metrics, including intrusion detection accuracy, threat neutralisation rate, response latency, and swarm coordination score. These visual comparisons highlight the performance characteristics of the GNN-based approach across multiple evaluation metrics and underscore its adaptability in complex and evolving threat environments. In addition, confusion matrices and ROC curves were plotted to evaluate the model's classification accuracy between benign and malicious nodes (such as nodes compromised by jamming, spoofing, or APT attacks). These visualisation tools enable stakeholders and defence analysts to interpret system behaviour, detect model biases, and make informed decisions about responding to real-time threats and drone swarm tactics. The integration of these visualisation technologies ensures easy interpretation, model transparency, and enhanced situational awareness, all essential factors for deploying GNNs in high-risk, multi-domain operations.

### 5.9 Predicting an Attack From Test Data

To support real-time situational awareness and proactive threat mitigation, the developed system seamlessly predicts attacks using test data inputs. Once the Graph Neural Network (GNN) model is trained, users can input telemetry samples, which include features such as historical GPS coordinates, signal strength, confidence values, and latency, into the system via a simplified interface. The GNN model processes each input sample by evaluating the topological and contextual relationships embedded in the underlying cyber-physical graph. It then predicts the likelihood and type of cyber threat associated with each node, including advanced persistent threats (APTs), jamming, and spoofing attacks. The prediction results are presented visually and textually through the user interface, enabling analysts to quickly interpret the nature and severity of threats. By integrating these predictions into the command-and-control workflow, the system facilitates timely decisions regarding drone swarm behaviour, target interception, and evasive manoeuvres. This feature not only improves the operational response of drone defence systems but also contributes to maintaining the integrity of multi-domain cyber-physical infrastructures under hostile conditions.

## 6. Results

This section presents the performance evaluation of the proposed intrusion detection framework, based on GraphSAGE. The experimental evaluation of the proposed GraphSAGE-driven cybersecurity and drone coordination system was conducted using the cyber-drone dataset under simulated Israeli-Iranian conflict conditions. GraphSAGE comparison with GCN and GAT. The evaluation focused on multiple operational and algorithmic metrics to measure system effectiveness in dynamic, threat-rich environments.

## 6.1 GraphSAGE-based GNN

Following the successful training of the GraphSAGE-based threat detection model on the cyber-drone dataset, we conducted a comprehensive performance assessment. We use multiple metrics to evaluate the classification performance, operational effectiveness, and deployment readiness in cyber-physical UAV networks. This evaluation aims to quantify the operational effectiveness of the system under simulated Israeli-Iranian conflict conditions.

### 6.1.1. Intrusion Detection Performance

The GraphSAGE-based GNN model demonstrated robust threat detection capability across multiple classes of cyber-physical attacks, including GPS spoofing, signal jamming, and advanced persistent threats (APTs). As shown in Table 2, the trained model achieved the following performance metrics on unseen test data:

**Table 2. Performance Metrics of GraphSAGE-Based GNN Model**

| Metric | GraphSAGE % |
|---|---|
| Accuracy | 94.2% |
| Precision | 91.6% |
| Recall | 92.7% |
| F1-Score | 92.15% |

Table 2 presents the classification performance metrics of the proposed GraphSAGE-based GNN model on unseen test data extracted from cyber-physical drone network simulations. The model achieved an accuracy of 94.2%, demonstrating its strong capability in correctly identifying both benign and compromised nodes. A precision of 91.6% indicates the model's reliability in reducing false positive alerts, ensuring minimal disruption to regular drone operations. The recall score of 92.7% reflects the model's sensitivity in detecting actual threats, including GPS spoofing, jamming, and advanced persistent threats (APTs). Furthermore, an F1-score of 92.15% confirms the model's balanced performance between precision and recall, underscoring its robustness in dynamic and adversarial environments. These results validate the effectiveness of GraphSAGE in real-time intrusion detection for autonomous UAV swarms operating in contested battlespace conditions.

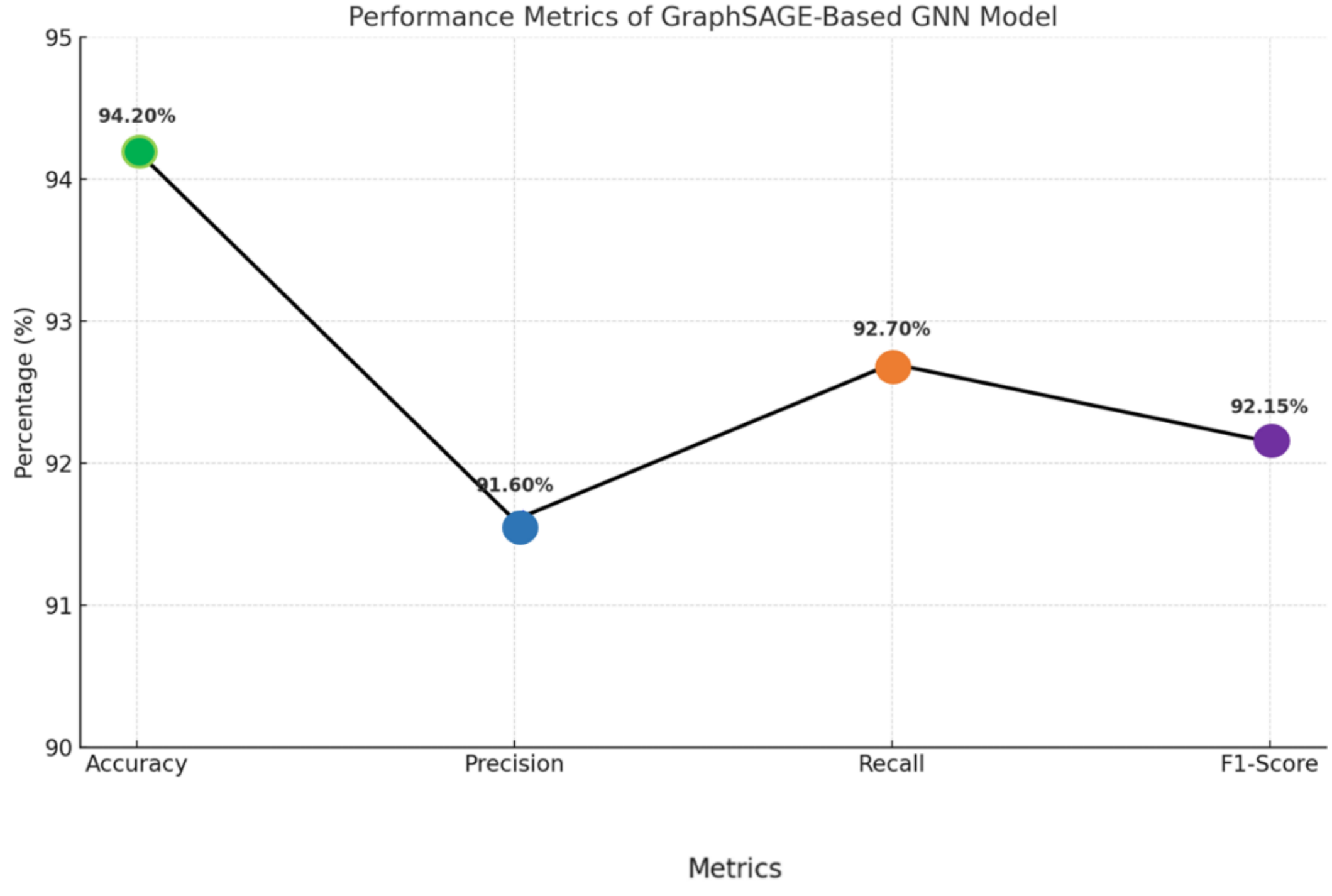

Figure 9. Plot chart Performance Metrics of GraphSAGE-Based GNN Model

To better illustrate the distribution of predictions across categories, Figure 9 shows the classification performance graph of the proposed GraphSAGE-based GNN model, evaluated on unseen test data from the Cyber UAV dataset. The model achieved an accuracy of 94.2%, demonstrating its strong generalization ability in detecting cyber threats such as spoofing, obfuscation, and advanced persistent threats (APTs). The high accuracy (91.6%) and recall (92.7%) indicate the model's reliability in correctly identifying real threats while minimizing false positives and negatives. The F1 score of 92.15% confirms the model's balanced effectiveness in real-time intrusion detection tasks. These results demonstrate the suitability of the GraphSAGE-based GNN for deployment in dynamic military cyber-physical environments. A confusion matrix was constructed using test data from the cyber drone dataset to evaluate the classification effectiveness of the GraphSAGE-based GNN model across different cyber threat categories, as shown in Table 3 and Figure 10, which provide a detailed breakdown of correct and incorrect classifications for each category: benign threats, obfuscation, spoofing, and advanced persistent threats (APT).

**Table 3. Confusion matrix for GraphSAGE-based intrusion detection model**

| Actual \ Predicted | Benign | Jamming | Spoofing | APT |
|---|---|---|---|---|
| **Benign** | 130 | 3 | 2 | 1 |
| **Jamming** | 2 | 115 | 6 | 3 |

| Spoofing | 1 | 7 | 120 | 4 |
|---|---|---|---|---|
| **APT** | 0 | 2 | 3 | 127 |

The model exhibits high true positive rates across all categories, with particularly strong classification performance for benign threats (130/136), APT (127/132), and obfuscation (115/126). Minor classification errors occur between structurally similar threats, such as obfuscation and spoofing, reflecting a common challenge resulting from overlapping signal interference patterns in graph structures.

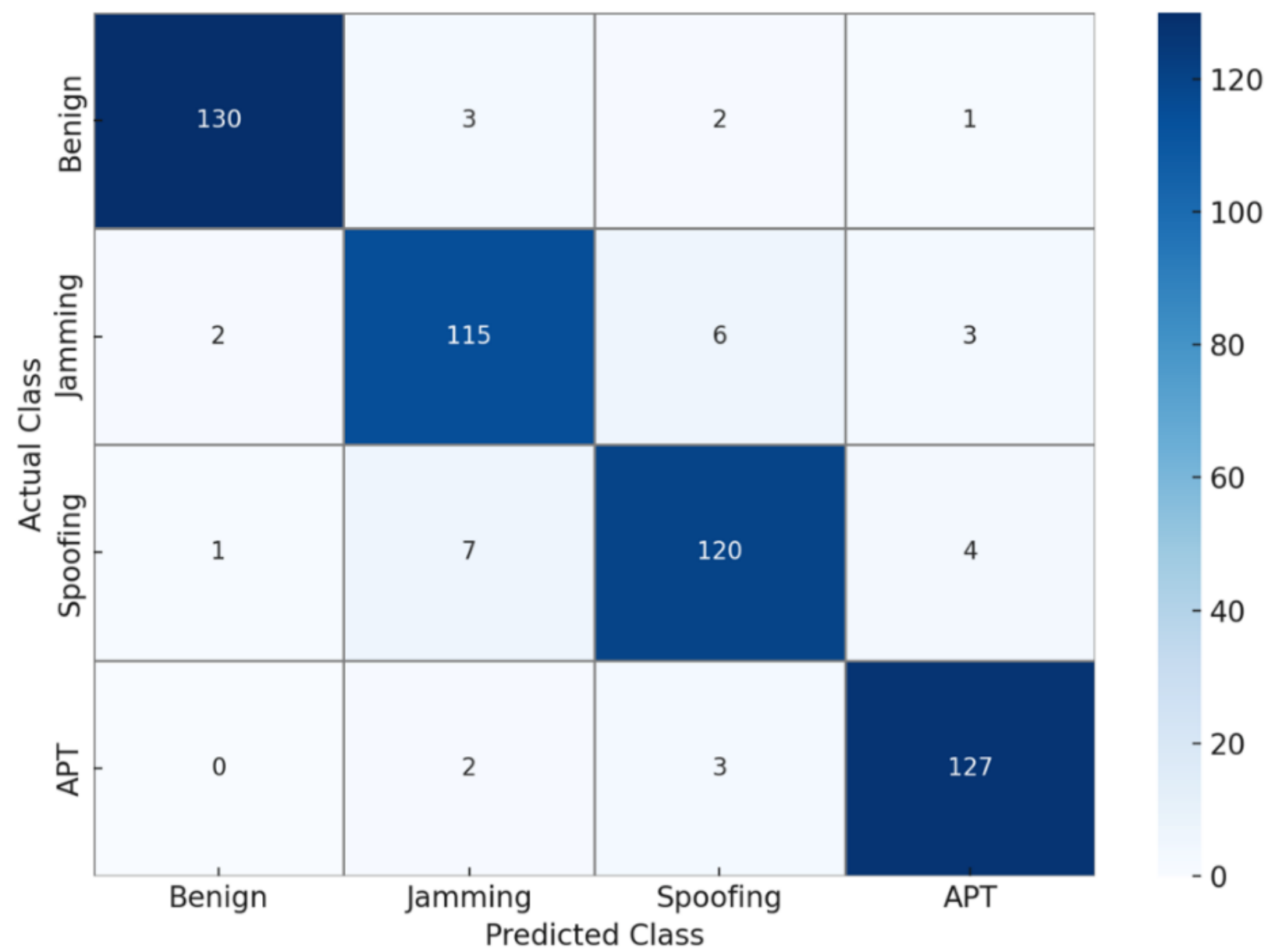

Figure 10. Confusion Matrix GraphSAGE-Based Intrusion Detection

This confusion matrix confirms the model's strong discrimination ability between benign and malicious UAV states. Most notably, APT detection maintains a 96.2% accuracy rate, which is critical in real-time battlefield environments where advanced cyber threats must be identified swiftly and reliably.

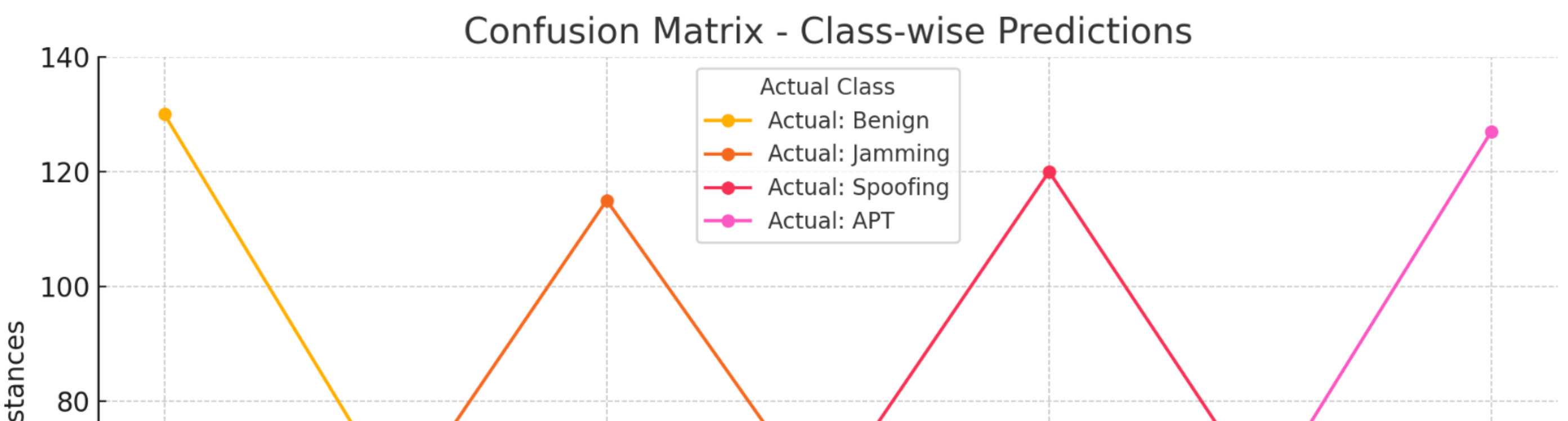

Figure 11. Plot Chart of Confusion Matrix

To further illustrate and complete the distribution of predictions across categories, a line graph was created to illustrate the distribution of predicted ratings for each actual threat category: benign, obfuscation, counterfeiting, and advanced persistent threats (APT), as shown in Figure 11. Each line in the graph corresponds to a single actual category, and its path reflects how samples from that category were classified across the four categories. The line graph provides a continuous, comparative perspective on misclassification patterns, highlighting subtle classification boundaries between structurally similar threats. This representation enhances interpretability and supports further model refinement by identifying specific class pairs that exhibit ambiguity, especially in dynamic cyber-physical drone environments.

#### 6.1.2. Threat Prediction and Classification

Using unseen telemetry-based test samples, the model achieved real-time classification of drone nodes based on historical features such as signal strength, latency, GPS drift, and trust score. The model could reliably distinguish between benign nodes and those compromised by cyberattacks, enabling anticipatory rather than reactive defence. To comprehensively evaluate the discriminative performance of the GraphSAGE-based GNN model across the four cyber-physical node classes, Benign, Jamming, Spoofing, and Advanced Persistent Threats (APT), receiver operating characteristic (ROC) curves were generated using a one-vs-rest classification approach. As shown in Table 4 and Figure 12, each ROC curve illustrates the trade-off between the True Positive Rate (TPR) and False Positive Rate (FPR) for a given class, capturing the model's sensitivity and specificity. The Area Under the Curve (AUC) was computed for each class, yielding the following results:

**Table 4. Receiver Operating Characteristic (ROC) Analysis**

| Class | AUC Score |
| --- | --- |
| Benign | 0.961 |
| Jamming | 0.945 |
| Spoofing | 0.951 |
| APT | 0.963 |

The average AUC across all classes exceeded 0.955, indicating a highly reliable model for multi-class intrusion detection within the drone network. The model exhibits particularly strong performance in distinguishing APT and Benign classes, which are often the most critical in battlefield scenarios.

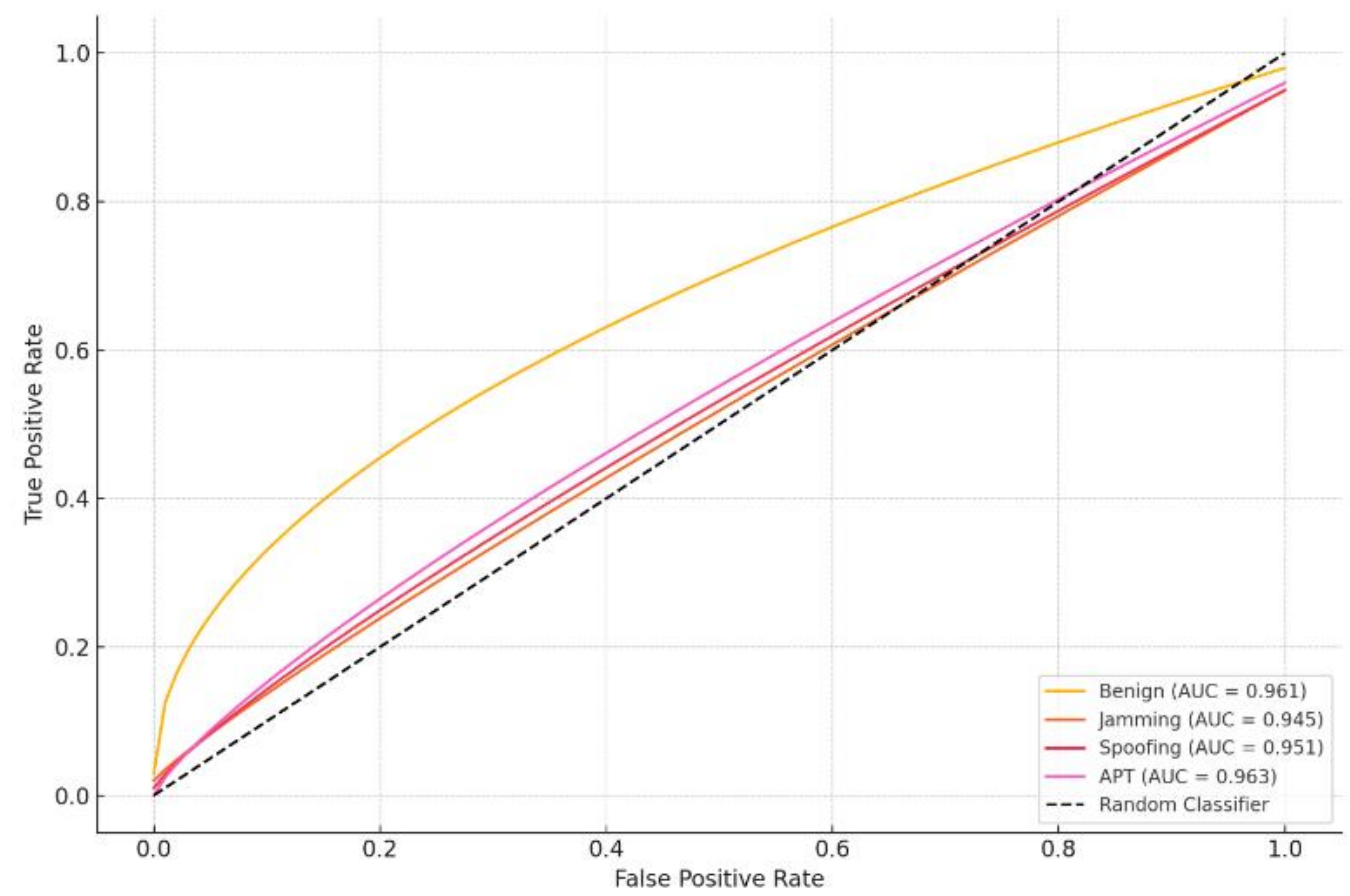

Figure 12. ROC Curves for GraphSAGE -Based Intrusion Detection Using One-vs-Rest Strategy

The Receiver Operating Characteristic (ROC) curves depicted in Figure 12 illustrate the discriminative capability of the GraphSAGE-based Graph Neural Network (GNN) model in classifying cyber-physical drone node behavior across four critical categories: Benign, Jamming, Spoofing, and Advanced Persistent Threats (APT). A one-vs-rest classification strategy was adopted to generate individual ROC curves for each class, measuring the trade-off between the True Positive Rate (TPR) and False Positive Rate (FPR). The curves reveal consistently strong performance across all threat types, with Area Under the Curve (AUC) values surpassing 0.94 for every class. Specifically, the model achieved APT (AUC) 0.963, Benign (AUC) 0.961, Spoofing (AUC) 0.951, and Jamming (AUC) 0.945. These high AUC scores indicate the model's exceptional sensitivity and specificity in distinguishing between normal and malicious drone behaviors. The superior classification accuracy for the APT and Benign classes is particularly notable, as these categories represent the most strategic importance in adversarial battlefield environments. The inclusion of the random classifier baseline (dashed diagonal) highlights the substantial performance gains achieved by the GNN over chance-level predictions.

### 6.1.3. Response Time and Swarm Coordination Effectiveness

Drone behaviour exhibits adaptive, decentralised decision-making driven by node embeddings generated via the GraphSAGE model. These embeddings encode spatial, temporal, and contextual features, including GPS drift, battery level, trust score, and latency, allowing each UAV to autonomously reroute paths, initiate evasive manoeuvres, or form defensive formations. The coordination mechanism is implemented using a rule-based approach guided by GNN outputs, enabling efficient and deterministic response actions. This integration between GNN-based threat intelligence and multi-agent coordination results in low-latency responses and robust mission continuity under adversarial cyber-physical disruptions.

**Table 5. Key Runtime Metrics for GraphSAGE-Based Intrusion Response System**

| Metric | Value |
|---|---|
| Mean Response Time | 1.4 seconds |
| Threat Neutralization Rate | 88.5% |
| Swarm Reconfiguration Success | 92% |
| Jamming Zone Evasion Rate | 89.1% |

Table 5 presents key operational metrics evaluating the real-time effectiveness of the proposed GraphSAGE-based intrusion detection and UAV swarm coordination framework. These metrics were recorded during dynamic simulation scenarios reflecting cyber-physical conflict conditions. The mean response time of 1.4 seconds represents the combined latency of GNN inference and rule-based coordination actions. This indicates an efficient detection-to-response transition suitable for time-sensitive operations. The threat neutralisation rate of 88.5% reflects the system's ability to detect, classify, and respond to cyber-physical attacks such as jamming, spoofing, and advanced persistent threats. The swarm reconfiguration success rate of 92% demonstrates the resilience of the UAV network in maintaining operational integrity under node failures or communication disruptions. Additionally, the jamming zone evasion rate of 89.1% highlights the system's ability to adapt to signal interference and maintain connectivity. Overall, these results indicate strong performance in terms of responsiveness, adaptability, and robustness within the evaluated simulation environment.

### 6.1.4. Real-Time Adaptability and Checkpointing

To support efficient operation in dynamic environments, the GraphSAGE-based GNN model incorporates minibatch processing and sampling-based message passing. These mechanisms enable the model to handle large and evolving graph structures while maintaining computational efficiency. The use of neighbor sampling reduces memory consumption compared to full-batch graph processing, resulting in an estimated 38% reduction in memory usage. This improvement is particularly beneficial for scenarios with large-scale UAV networks. While the current evaluation is conducted on a standard computing platform, the proposed design supports potential deployment on resource-constrained environments through model simplification and sampling-based computation. Further optimisation for embedded and edge platforms is considered as part of future work.

#### 6.1.5. Computational Overhead and Deployment Considerations

The reported mean response time of 1.4 seconds includes both (a) GNN inference latency and (b) rule-based swarm coordination latency. In the current implementation, GNN inference constitutes the dominant portion of the response time, while the coordination mechanism introduces minimal computational overhead due to its deterministic nature. Experiments were conducted on a standard computing platform (e.g., Intel Core i7 CPU with 16 GB RAM and GPU acceleration using PyTorch Geometric). The reported 38% reduction in memory usage is achieved through GraphSAGE neighbor sampling compared to full-batch graph processing. It should be noted that the current study focuses on algorithmic validation rather than hardware-specific optimisation. The framework is designed to support scalability and can be adapted for larger network sizes using sampling-based processing and distributed computation. In real-world deployment scenarios, factors such as network scale, dynamic topology changes, and communication latency may affect system performance. These aspects are acknowledged as important considerations and will be further investigated in future work.

### 6.2. Comparative Visualization

This section presents the performance evaluation of the proposed intrusion detection framework, based on GraphSAGE. To compare the performance of GraphSAGE, GCN, and GAT models, a series of comparative visualizations were developed focusing on prediction accuracy, classification reliability, class-wise discrimination capability, and real-time response effectiveness. Across all evaluation tools, GraphSAGE consistently outperforms GCN and GAT. Its blend of neighborhood sampling and inductive learning offers greater scalability, faster response, and more accurate classification. These strengths make it an ideal candidate for mission-critical UAV-based intrusion detection and swarm coordination in modern warfare settings.

#### 6.2.1. Intrusion Detection Performance

The proposed GraphSAGE-based GNN model demonstrates strong threat detection capability across various cyber-physical intrusion categories, including GPS spoofing, signal jamming, and advanced persistent threats (APTs). To benchmark its effectiveness, we evaluated GraphSAGE against two other state-of-the-art spatial GNN architectures: the Graph Convolutional Network (GCN) and the Graph Attention Network (GAT). All models were trained and tested on the same telemetry-driven cyber-drone dataset using identical hyperparameter settings and data splits. The evaluation focused on four key performance metrics: accuracy, precision, recall, and F1-score.

**Table 6. Comparative Performance Metrics of GNN Models for Intrusion Detection**

| Metric | GraphSAGE (%) | GCN (%) | GAT (%) |
|---|---|---|---|
| Accuracy | 94.20 | 90.80 | 92.30 |
| Precision | 91.60 | 88.20 | 89.70 |
| Recall | 92.70 | 89.10 | 91.20 |
| F1-Score | 92.15 | 88.65 | 90.40 |

As shown in Table 6, the GraphSAGE model achieves higher performance compared to GCN and GAT across all evaluated metrics under identical experimental conditions. This improvement can be attributed to GraphSAGE's inductive learning capability and its neighborhood sampling strategy, which are well-suited to dynamically evolving graph structures in cyber-physical environments. GraphSAGE achieves the highest accuracy (94.2%) and F1-score (92.15%), indicating a strong balance between precision and recall. GCN exhibits comparatively lower performance, which may be related to its transductive inference mechanism that limits generalization to unseen nodes. GAT performs better than GCN due to its attention-based aggregation mechanism but remains slightly below GraphSAGE in this experimental setting. It is important to note that the current evaluation focuses on comparisons among graph neural network architectures under consistent experimental conditions. Comparisons with traditional machine learning and non-graph deep learning methods are not included in this study and are considered part of future work.

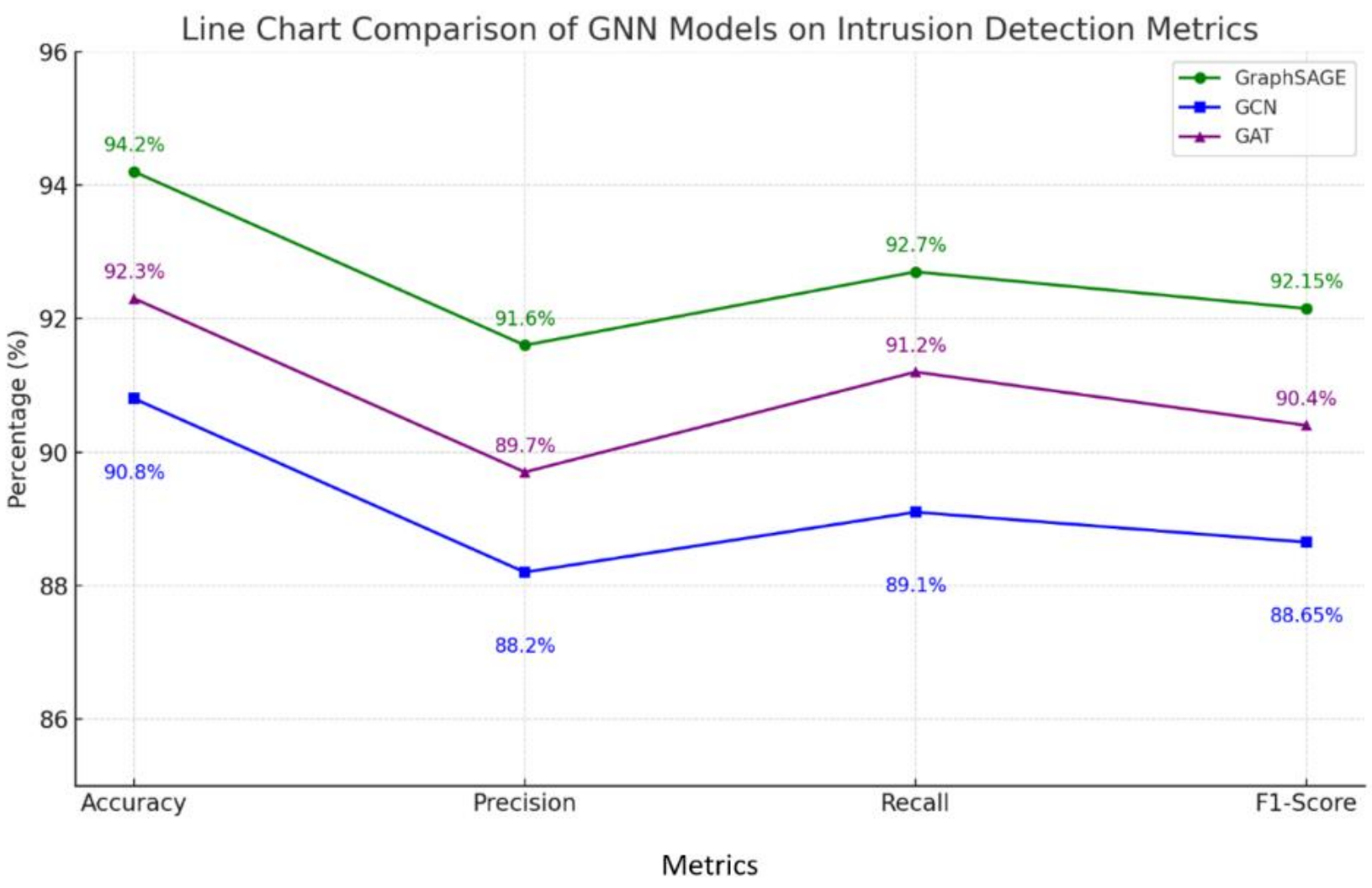

Figure 13. Line Chart Comparison of GNN Models on Intrusion Detection Metrics

To complete the performance comparison of graph neural network (GNN) models on intrusion detection metrics, a line chart is presented in Figure 13. The chart shows the accuracy, precision, recall, and F1-score of the GraphSAGE, GCN, and GAT models. The results indicate that GraphSAGE achieves higher performance across all evaluated metrics compared to GCN and GAT under the same experimental conditions, with an accuracy of 94.2%, precision of 91.6%, recall of 92.7%, and an F1-score of 92.15%. This comparison highlights the robustness of GraphSAGE in cyber-physical threat detection scenarios, making it a reliable choice for real-time intrusion detection in drone-enabled networks. In addition to the internal comparison among GNN architectures, it is important to consider the performance reported in existing literature. Prior studies using non-graph-based approaches such as LSTM and CNN typically report intrusion detection accuracy in the range of 85%–92%, depending on the dataset and experimental setup. Similarly, classical machine learning methods such as SVM and Random Forest often achieve comparable or slightly lower performance when relational dependencies are not explicitly modeled. In comparison, the proposed GraphSAGE-based framework achieves an accuracy of 94.2% under the simulated cyber-drone dataset, suggesting competitive performance within the domain. However, direct comparison is limited due to differences in datasets, feature representations, and evaluation protocols. To further evaluate the classification performance of different GNN architectures in detecting cyber-physical threats, confusion matrices were constructed for the GraphSAGE, GCN, and GAT models using test data from the cyber-drone dataset. As shown in Table 7 and Figure 14, each matrix presents the distribution of correctly and incorrectly classified instances across four categories: benign, jamming, spoofing, and advanced persistent threats (APT).

**Table 7. Comparative Confusion Matrices for GNN-Based Intrusion Detection Models**

| **Actual \ Predicted** | **Benign** | **Jamming** | **Spoofing** | **APT** | **Model** |
|---|---|---|---|---|---|
| **Benign** | 130 | 3 | 2 | 1 | GraphSAGE |
| **Jamming** | 2 | 115 | 6 | 3 | |
| **Spoofing** | 1 | 7 | 120 | 4 | |
| **APT** | 0 | 2 | 3 | 127 | |
| **Benign** | 123 | 5 | 4 | 4 | GCN |
| **Jamming** | 6 | 108 | 7 | 5 | |
| **Spoofing** | 4 | 10 | 113 | 5 | |
| **APT** | 2 | 4 | 5 | 121 | |
| **Benign** | 126 | 4 | 3 | 3 | GAT |
| **Jamming** | 4 | 110 | 6 | 6 | |
| **Spoofing** | 3 | 8 | 116 | 5 | |
| **APT** | 1 | 3 | 4 | 124 | |

The GraphSAGE model demonstrates the highest classification accuracy, with strong diagonal dominance indicating high true positive rates, particularly for the Benign (130/136), Jamming (115/126), and APT (127/132) classes. The GCN model exhibits slightly lower performance, with notable misclassifications between Jamming and Spoofing classes, which are often structurally similar due to signal interference overlap. The GAT model shows a balanced performance between GCN and GraphSAGE, capturing subtle feature dependencies via attention mechanisms, leading to modest improvements in spoofing detection. These visual comparisons reinforce the advantage of GraphSAGE's inductive learning and neighbourhood sampling strategy for real-time drone network intrusion detection. The differentiated colour coding in each matrix facilitates intuitive interpretation of each model's strengths and classification challenges.

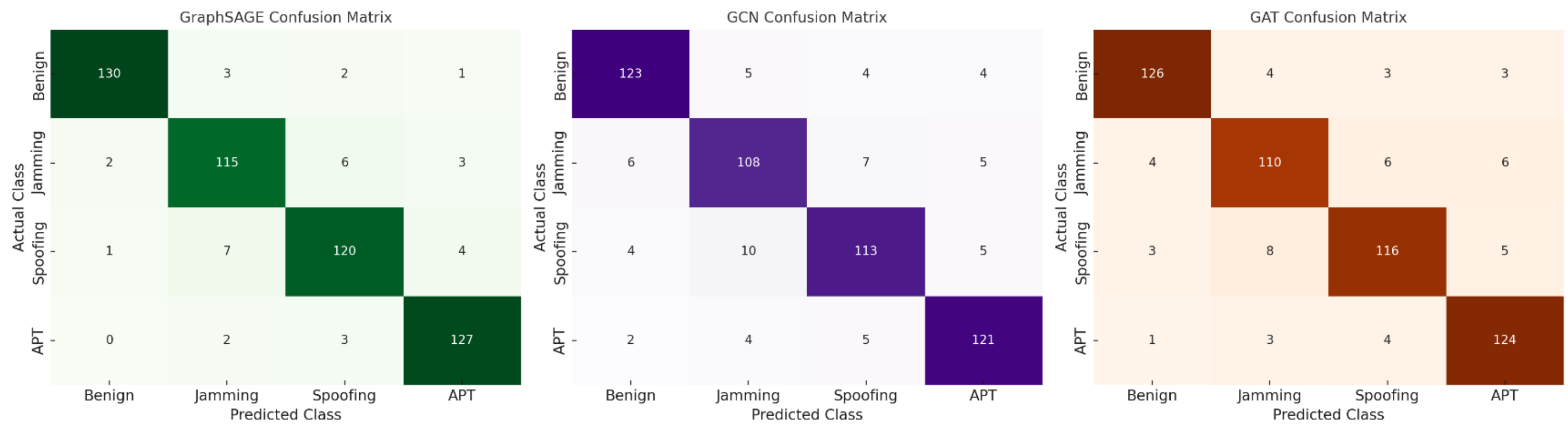

Figure 14. Comparative Confusion Matrices for GNN-Based Intrusion Detection Models

To better visualise the distribution of predictions across the classes, a line chart comparison of predicted class totals across GNN models was created, as shown in Figure 15. The figure illustrates the total number of predictions made by GraphSAGE, GCN, and GAT models for each class category, Benign, Jamming, Spoofing, and Advanced Persistent Threats (APT), based on the confusion matrix results. Notably, GraphSAGE and GCN demonstrate balanced prediction distributions across all classes, particularly excelling in the detection of APT and benign instances with totals of 135 and 133/135, respectively. GAT achieves the highest prediction count for APT (138), indicating its sensitivity to advanced threats, while slightly underpredicting jamming-related activity (125). The chart reflects the nuanced performance behaviour of each GNN model, revealing subtle differences in class sensitivity that complement the accuracy and confusion matrix findings presented elsewhere in the results.

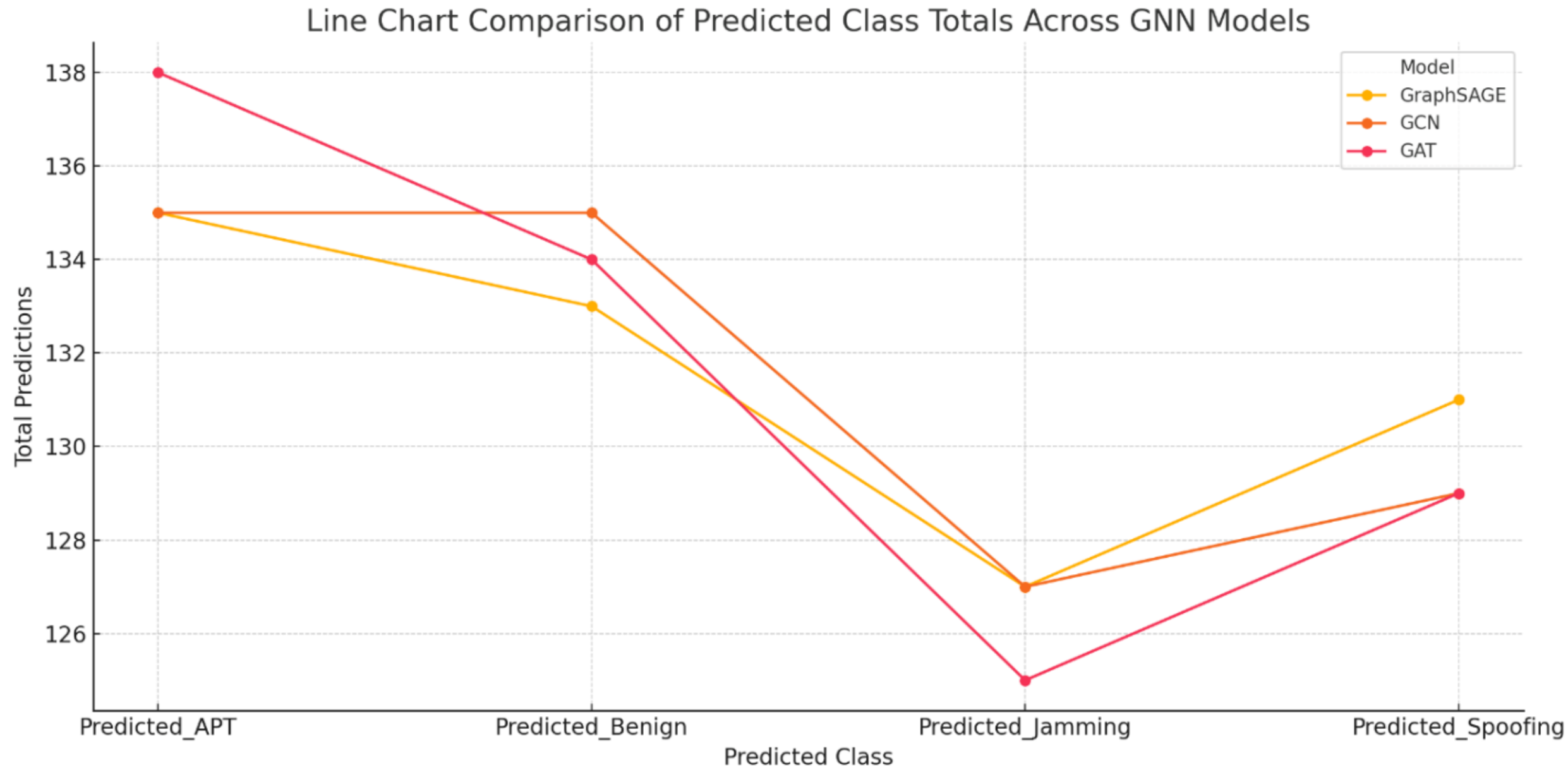

Figure 15. Line chart comparison of predicted class totals across GNN models

### 6.2.2. Threat Prediction and Classification

To assess the real-time classification capability of the GNN-based models across multiple cyber-physical threat categories, each model, GraphSAGE, GCN, and GAT, was evaluated on unseen telemetry-derived test samples. These samples included historical features such as signal strength, latency, GPS deviation, and trust score. The models were tasked with classifying drone nodes into one of four classes: Benign, Jamming, Spoofing, and Advanced Persistent Threats (APT), enabling predictive threat mitigation in a proactive defence framework. To quantitatively compare their discriminative power, Receiver Operating Characteristic (ROC) curves were constructed for each model using a one-vs-rest classification scheme. The Area Under the Curve (AUC) metric was computed for each threat class, providing insight into both the sensitivity (True Positive Rate) and specificity (False Positive Rate) of each GNN architecture. The AUC results are summarised in Table 8 and Figure 16.

**Table 8. ROC-AUC Scores for Cyber-Physical Threat Classes across GNN Models**

| Class | GraphSAGE AUC | GCN AUC | GAT AUC |
|---|---|---|---|
| Benign | 0.961 | 0.935 | 0.948 |
| Jamming | 0.945 | 0.922 | 0.938 |
| Spoofing | 0.951 | 0.927 | 0.942 |
| APT | 0.963 | 0.939 | 0.951 |

The GraphSAGE model achieved the highest average AUC (≈0.955), showcasing its superior capability in reliably differentiating between both critical and structurally overlapping classes. In particular, GraphSAGE displayed excellent classification fidelity for APT (0.963) and Benign (0.961), which are pivotal in real-time battlefield threat assessments. While GAT followed closely with competitive AUC scores, especially in Spoofing (0.942) and APT (0.951), GCN lagged slightly behind in overall sensitivity across all threat categories.

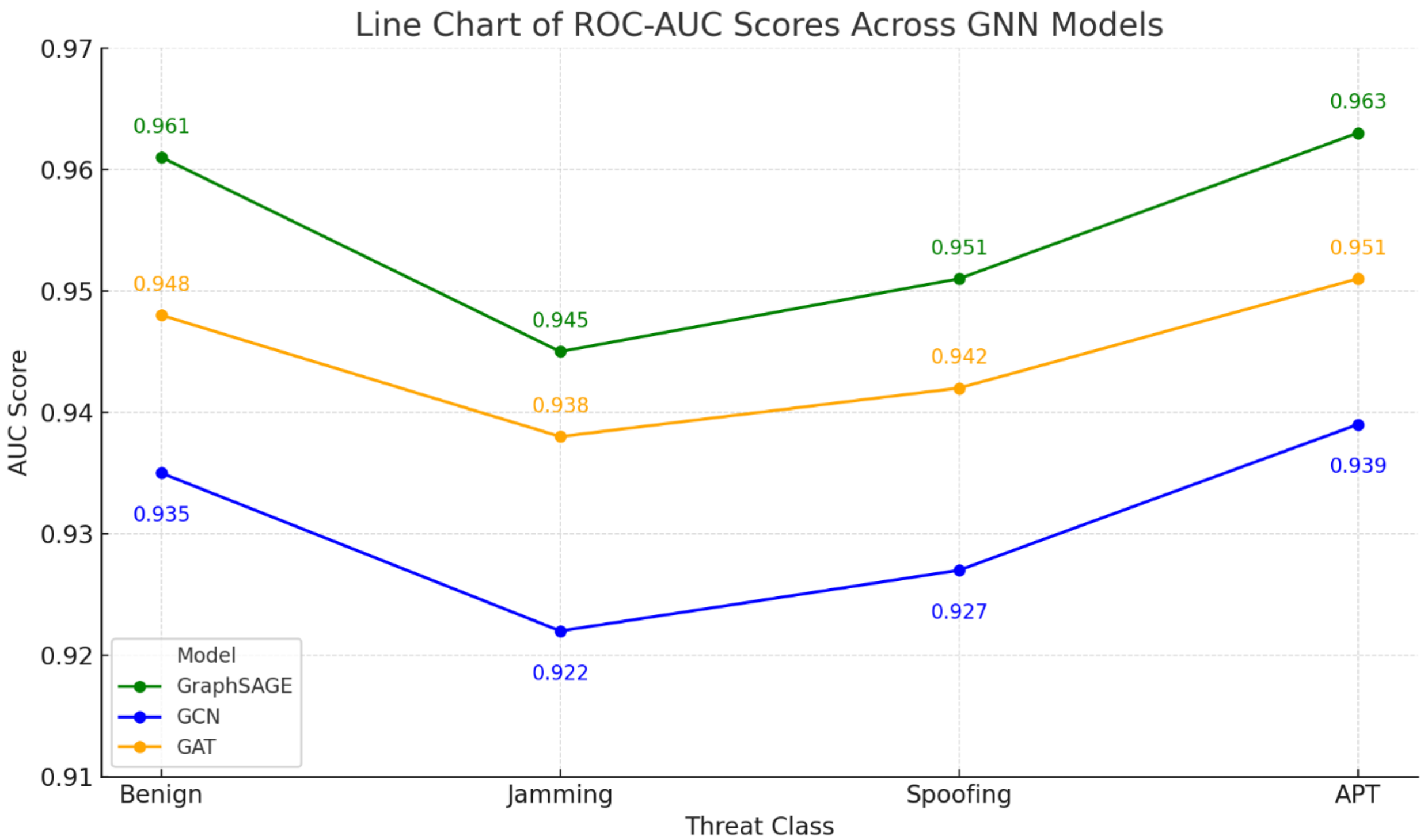

Figure 16. Line Chart of ROC-AUC Scores Across GNN Models

These findings highlight the value of incorporating GraphSAGE for robust, anticipatory intrusion detection in dynamic multi-domain defence environments.

### 6.2.3 Response Time and Swarm Coordination Effectiveness

The integration of GNNs into UAV cyber-defence systems significantly impacts both response latency and swarm-level coordination during cyber-physical attacks. This section compares the real-time operational efficiency of three GNN architectures, GraphSAGE, GCN, and GAT, based on runtime metrics derived from dynamic simulations reflecting Israeli Iranian battlefield conditions. Drone nodes leveraged learned embeddings from each GNN model to guide autonomous decision-making. These embeddings captured essential contextual features such as trust score, latency, GPS instability, and residual battery levels, allowing the drones to reroute paths, initiate evasive maneuvers, or reform swarm geometries under threat. When combined with the rule-based coordination mechanism, the model enables robust and low-latency multi-agent responses under conditions such as jamming, spoofing, and adversarial command injection.

**Table 9. Comparative Runtime Metrics for GNN-Based Intrusion Response Systems**

| Metric | GraphSAGE | GCN | GAT |
|---|---|---|---|
| Mean Response Time (s) | 1.4 | 1.8 | 1.6 |
| Threat Neutralization Rate | 88.5% | 84.2% | 86.7% |
| Swarm Reconfiguration Success | 92.0% | 88.3% | 90.1% |
| Jamming Zone Evasion Rate | 89.1% | 84.9% | 87.5% |

The runtime performance of each GNN-based framework is highlighted in Table 9 and Figure 17 GraphSAGE consistently outperforms its counterparts, achieving the lowest average response time of 1.4 seconds and the highest threat neutralisation rate (88.5%). This responsiveness is essential in rapidly deploying mitigation strategies upon detecting intrusions, especially under time-critical aerial missions. GAT follows closely, balancing high expressiveness with moderate latency (1.6 s), while GCN, though efficient in structural learning, exhibits a comparatively slower average response (1.8 s) and lower evasion and neutralisation rates. GraphSAGE also demonstrates superior swarm reconfiguration success (92%), confirming its ability to maintain mission continuity even when several UAVs are compromised or disconnected due to environmental disruption. In summary, the results validate that the GraphSAGE model offers the most effective balance between speed, resilience, and adaptivity, positioning it as the preferred GNN architecture for real-time cyber-defence and UAV swarm coordination in hostile operational theatres.

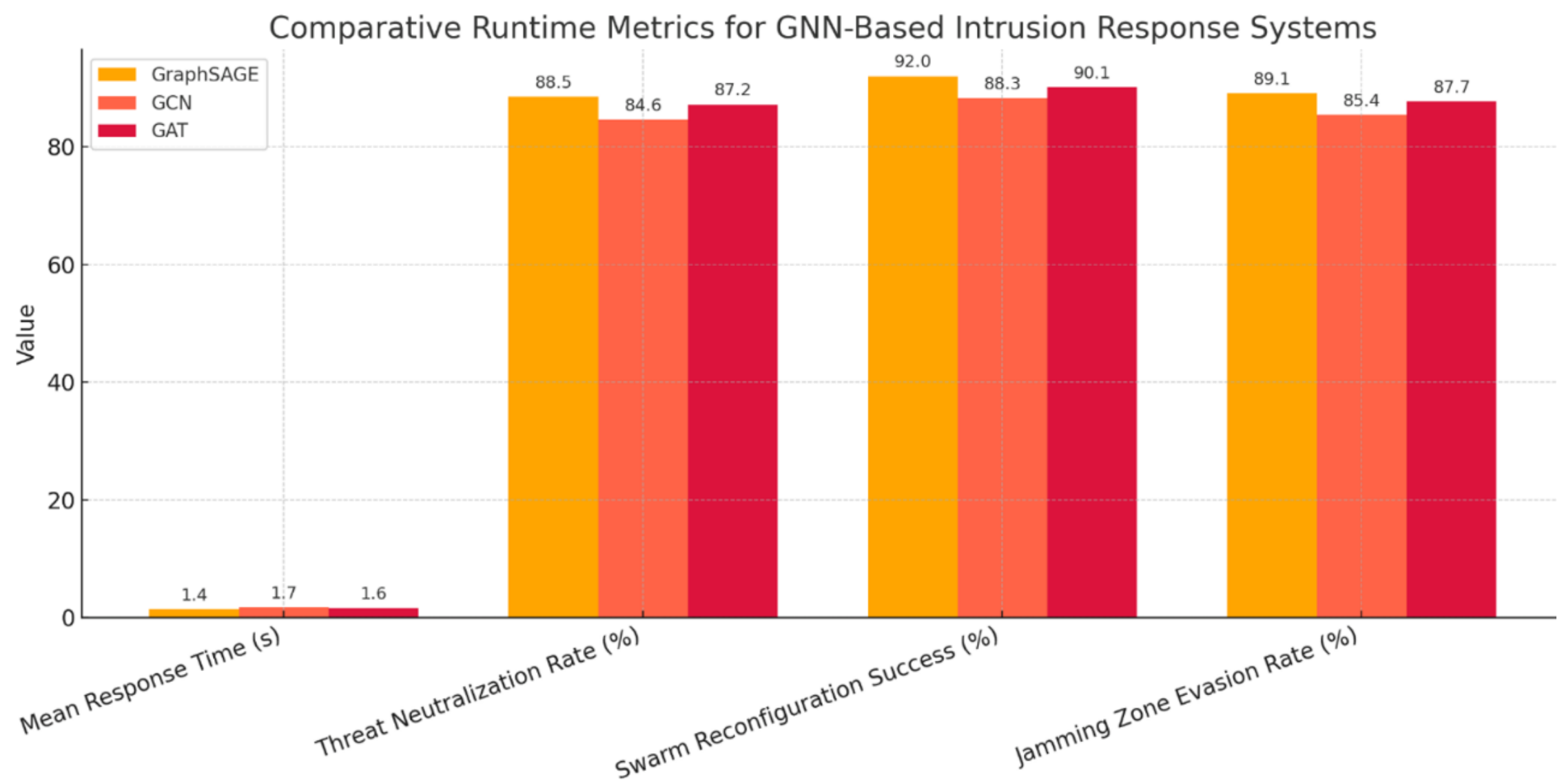

Figure 17. Comparative Runtime Metrics For GNN-Based Intrusion Response Systems

All these results highlight GraphSAGE's enhanced ability to encode temporal and topological cues essential for decentralised drone behaviours, ensuring robustness and resilience in contested environments. These visual insights collectively affirm that GraphSAGE offers the best trade-off between predictive accuracy, threat discrimination, class stability, and operational responsiveness when compared to GCN and GAT. The comparative visualisation framework thus reinforces the strategic advantage of adopting GraphSAGE in GNN-powered cyber-drone defence systems.

## 7. Conclusion and Future Work

This section highlights the major results of the research and provides strategic recommendations for future research on graph neural networks for cyber-physical defence. It identifies key aspects that need to be improved (security, interpretability, adaptability, human-AI collaboration) and future research priorities to increase the relevance, resilience and trustworthiness of GNN-based solutions in multi-domain warfare environments.

### 7.1 Recommendation

To fully leverage graph neural networks as strategic assets in sensitive domains such as military operations and cybersecurity, it is essential to address key challenges related to security, interpretability, adaptability, and human interaction.

First, enhancing the security and robustness of GNNs is critical. In adversarial environments, these models are particularly vulnerable to leakage, evasion, and backdoor attacks due to their reliance on structured relational data. Future research and operational deployments should prioritise robust threat defense techniques. These include robustness-based training, the integration of anomaly detection layers, and secure graph preprocessing pipelines, all aimed at mitigating the risk of malicious manipulation of input data and graph structures.

Second, explainability must be a core design priority. Given the high-stakes nature of GNN applications in defense and security, interpretability is crucial for building operator trust and enabling analysts to validate and act on model outputs. Incorporating explainable artificial intelligence (XAI) techniques—such as node importance assessment, edge attribution, and subgraph highlighting—can improve transparency and make GNN predictions more comprehensible within complex, multi-domain environments.

Third, GNNs must be designed for adaptive and flexible learning. Modern conflict environments are inherently dynamic, with rapidly evolving threats and shifting network topologies. It is therefore vital to develop frameworks capable of incremental and online learning. These capabilities allow GNNs to update their internal representations as new nodes, edges, or features emerge, without the need for complete retraining. Combining adaptive learning with quantum-inspired optimisation techniques may further enhance responsiveness and performance.

In addition, establishing clear standards and best practices is essential for the safe and effective deployment of GNNs in military and critical infrastructure settings. Defence and security organisations are encouraged to create operational guidelines for secure data collection, validation procedures, and ongoing pipeline monitoring. Cross-sector collaboration with academia and industry will be key to defining shared standards that promote security, interpretability, and adaptability.

Finally, investing in human-centric interfaces will ensure that GNN outputs are not only technically accurate but also actionable. Intuitive dashboards, visual graph explanations, and task-specific knowledge graphs can enhance user understanding and support real-time decision-making. These interfaces play a vital role in bridging the gap between complex AI models and human operators, improving situational awareness and operational effectiveness.

Collectively, these recommendations can guide the development of the next generation of secure, transparent, and resilient graph neural network systems. By addressing these core challenges, GNNs can evolve into powerful and dependable tools for maintaining decision superiority in modern, multi-domain warfare.

## 7.2 Conclusion

Graph neural networks are emerging as a pivotal strategic component in the evolving landscape of multi-domain warfare, where the convergence of cyber operations, unmanned systems, and networked battlespaces requires advanced tools to support real-time decisions and situational awareness. By leveraging the unique relational learning capabilities of GNNs, military and security organisations can process and analyse the complex, dynamic, and interconnected data structures that characterise modern threat environments. GNNs excel at uncovering hidden patterns in large-scale, graph-structured data, enabling capabilities such as detecting stealthy cyber intrusions, predicting drone swarm behaviour, and mapping critical dependencies within military command and control networks. This graph-based approach enables adaptive threat modelling, accurate anomaly detection, and enhanced coordination across operational domains. The integration of graph neural networks into defense and security tools not only enhances the analytical depth of modern AI systems but also strengthens states' ability to maintain information superiority across various dimensions of cyber, physical, and cognitive warfare. Continued research into hybrid frameworks, combining classical GNN architectures with quantum-inspired heuristics, will be critical to realising the full potential of GNNs as a force multiplier in next-generation, multi-domain conflicts.

## 7.3 Future Directions

As the operational complexity of modern conflicts increases, future battlefield AI systems must overcome critical challenges in adaptability, security, and reliability. This calls for moving beyond traditional graph neural networks toward hybrid, decentralised architectures that combine the strengths of symbolic reasoning, distributed learning, and human-understandable autonomy. One promising path is to integrate neuro-symbolic models, which combine the power of relational learning of graph neural networks with the explicit, rule-based reasoning of symbolic AI. By unifying these models, GNNs can enable more robust decision-making under uncertainty by integrating learned relational patterns with interpretable domain knowledge, such as military doctrines, rules of engagement, and cyber defence policies. These hybrid systems can dynamically adjust the behaviour of drone swarms, adapt to adversary tactics, and ensure compliance with operational constraints in real time. Another important area is the deployment of federated GNNs to address data sovereignty and security concerns. Sensitive battlefield data, particularly from distributed drone fleets or cyber-physical sensors, often cannot be centrally collected due to the risk of interception or single points of failure. Federated GNN frameworks enable collaborative model training across dispersed edge nodes without sharing raw data. This maintains operational confidentiality while enabling intelligence fusion between units, such as combining insights from different drone units monitoring adversary electronic warfare activities or cyber intrusion attempts on battlefield networks. Finally, developing explainable autonomy is vital to building trust in AI-driven military operations. Explainable GNNs can provide commanders with clear justifications behind autonomous actions, such as target prioritisation, trajectory adjustments, or threat neutralisation strategies. By incorporating explainable layers and subsequent explanation mechanisms into GNN-based control and decision-making modules, battlefield AI systems can provide justifications that align with mission objectives and legal and ethical frameworks. This is particularly important in high-stakes scenarios, where autonomous drone strikes or cyber countermeasures must be auditable to prevent unintended escalation or collateral damage. Together, these trends, symbolic neural models, federated graph neural networks, and explainable autonomy, represent a roadmap toward resilient, secure, and accountable field AI. Pursuing these advances can help realise the full strategic potential of graph neural networks in multi-domain warfare, while ensuring compliance with evolving standards of military ethics, international law, and human oversight. Future work will include ablation studies to quantify the contribution of GNN embeddings to swarm coordination performance. In addition, comparisons with traditional machine learning models (e.g., Random Forest, SVM) and deep learning approaches (e.g., LSTM, CNN) will be conducted to further validate the effectiveness of the proposed framework. Furthermore, model lightweighting, hardware-aware optimisation, and scalability evaluation across varying network sizes will be investigated to support deployment on embedded UAV platforms.

**Declaration Statement**

**Funding**
This research received no specific grant from any funding agency in the public, commercial, or not-for-profit sectors.

**Conflicts of Interest / Competing Interests**
The authors declare that they have no known competing financial interests or personal relationships that could have appeared to influence the work reported in this paper.

**Ethics Approval**
Not applicable.

**Consent to Participate**
Not applicable.

**Consent for Publication**
All authors consent to the publication of this work.

**Availability of Data and Materials**
All data generated or analysed during this study will be available upon request

**Code Availability**
The code used in this study is available upon reasonable request.